\documentclass[preprint2]{aastex}
\usepackage{psfig}

\newcommand{\kms}{\ifmmode{{\rm ~km~s}^{-1}}\else{~km~s$^{-1}$}\fi}

\newbox\grsign \setbox\grsign=\hbox{$>$} \newdimen\grdimen 
\grdimen=\ht\grsign
\newbox\laxbox \newbox\gaxbox
\setbox\gaxbox=\hbox{\raise.5ex\hbox{$>$}\llap
     {\lower.5ex\hbox{$\sim$}}}\ht1=\grdimen\dp1=0pt
\setbox\laxbox=\hbox{\raise.5ex\hbox{$<$}\llap
     {\lower.5ex\hbox{$\sim$}}}\ht2=\grdimen\dp2=0pt
\newcommand{\simgreat}{\mathrel{\copy\gaxbox}}
\newcommand{\simless}{\mathrel{\copy\laxbox}}

\slugcomment{to submit to {\it The Astrophysical Journal} }

\shorttitle{Radio Properties of IRAS Galaxies}
\shortauthors{Yun, Reddy \& Condon}

\begin{document}

\title{Radio Properties of Infrared Selected Galaxies \\
in the IRAS 2~Jy Sample}

\author{Min S. Yun\altaffilmark{1} and Naveen A. Reddy\altaffilmark{2}}
\affil{National Radio Astronomy Observatory,
	P.O. Box 0, Socorro, NM~~87801}
\and
\author{J. J. Condon}
\affil{National Radio Astronomy Observatory, 520 Edgemont Road, 
Charlottesville, VA 22903}

\altaffiltext{1}{present address: University of Massachusetts, Astronomy Department, Amherst, MA 01003 (e-mail: myun@astro.umass.edu)}

\altaffiltext{2}{present address: California Institute of Technology, Astronomy Department, Pasadena, CA 91125 (e-mail: nar@astro.caltech.edu)}

\begin{abstract}

The radio counterparts to the IRAS Redshift Survey galaxies are 
identified in the NRAO VLA Sky Survey (NVSS) catalog.  Our new catalog
of the infrared flux-limited ($S_{60\mu m}\ge2$ Jy) complete sample of
1809 galaxies lists accurate radio positions, redshifts, and 1.4 GHz
radio and IRAS fluxes.  This sample is six times larger in size and
five times deeper in redshift coverage (to $z\approx 0.15$) compared
with those used in earlier studies of the radio and far-infrared (FIR)
properties of galaxies in the local volume.  The well known radio-FIR
correlation is obeyed by the overwhelming majority ($\ge 98\%$) of the
infrared-selected galaxies, and the radio AGNs identified by their
excess radio emission constitute only about 1\% of the sample,
{\it independent of the IR luminosity}.  These FIR-selected galaxies
can account for the entire population of late-type field galaxies in the local
volume, and their radio continuum may be used directly to infer the
extinction-free star formation rate in most cases.
Both the 1.4 GHz radio and 60 $\mu$m infrared luminosity functions are
reasonably well described by linear sums of two
Schechter functions, one representing normal, late-type field galaxies
and the second representing starbursts and other luminous infrared
galaxies.  The integrated FIR luminosity density for the local volume
is $(4.8\pm0.5) \times 10^7 L_\odot$ Mpc$^{-3}$, less than 10\% of
which is contributed by the luminous infrared galaxies with
$L_{FIR}\ge 10^{11} L_\odot$.  The inferred {\it extinction-free}
star formation density for the local volume is 
$0.015\pm0.005~M_\odot$ yr$^{-1}$ Mpc$^{-3}$.

\end{abstract}

\keywords{ galaxies: luminosity function 
-- galaxies: starburst -- galaxies: active -- infrared: galaxies -- radio continuum: galaxies --- surveys}

\section{Introduction}

The Infrared Astronomical Satellite (IRAS; Neugebauer et al. 1984) 
was the first
telescope with sufficient sensitivity to detect a large number of
extragalactic sources at mid- and far-infrared wavelengths.  The
IRAS survey covering 96\% of the sky detected a sample of $\approx 20,000$
galaxies complete to 0.5 Jy\footnote{1 Jansky (Jy) $= 10^{-26}$ W
Hz$^{-1}$ m$^{-2}$.} in the 60 $\mu$m band, the majority of which had
not been previously cataloged.  While most of these are late-type
galaxies with modest infrared luminosity ($L_{IR}/L_B=0.1-5$), some
extremely bright galaxies emitting most of their light in the infrared
wavelengths were also detected ($L_{IR}\ge
10^{12}~L_\odot,~L_{IR}/L_B\ge50$; de Jong et al. 1984, Soifer et
al. 1984, 1987 -- see review by Sanders \& Mirabel 1996).  Their observed FIR
luminosity can reasonably be explained by ongoing star formation and
starburst activity, but some of the most luminous infrared galaxies
may host an active galactic nucleus (AGN).

There are several important reasons for investigating the radio
properties of FIR-emitting galaxies.  First of all, the observed radio
emission may be used as a direct probe of the very recent star forming
activity in normal and starburst galaxies (see a review by Condon
1992).  Nearly all of the radio emission at wavelengths longer
than a few cm from such galaxies is
synchrotron radiation from relativistic electrons and free-free
emission from H~II regions.  Only massive stars ($M\simgreat
8~M_\odot$) produce the Type II and Type~Ib supernovae whose remnants
(SNRs) are thought to accelerate most of the relativistic electrons,
and the same massive stars also dominate the ionization in H~II
regions.  This results in an extremely good correlation between radio
and far-infrared emission among galaxies (Harwit \& Pacini 1975, Condon et
al. 1991a,b).  While the use of radio continuum as a star-formation
tracer has been suggested before, the validity and details of the
method has not yet been quantified fully.  By examining the radio
properties of a large sample of FIR-selected galaxies, we plan to
address the utility of radio continuum emission as a probe of
star-forming activity in detail, including its limitations and the
magnitude of systematic errors.

Another important reason for examining their radio properties is to
investigate the frequency and energetic importance of AGNs among
FIR-bright galaxies.  If a luminous AGN is present, then the
associated radio continuum, in excess of the level expected from the
star formation, may be detected.  The presence of a moderately compact
nuclear radio continuum source is not a sufficient proof that an AGN
is the source of the observed luminosity.  On the other hand, absence
of significant radio emission associated with AGN activity may
indicate that AGNs are not energetically important, at least among the
FIR-selected galaxies (e.g. Crawford et al. 1996).

This radio study of the FIR-selected galaxies is particularly timely
and relevant in light of the recent exciting discovery of highly
obscured and luminous submm sources at cosmological distances
\citep{Smail97,Hughes98,Barger99,Eales99}.  Though optically
unremarkable, these dusty galaxies may dominate the star
formation and metal production at $z\ge2$.  The majority of these
submm galaxies may be too faint optically to yield their redshifts
even with the largest telescopes, and deep radio imaging observations
such as with the VLA may be the only way of identifying and inferring
their redshift distribution and star formation rate for now (Carilli
\& Yun 1999, Smail et al. 2000).  Therefore, understanding the radio
properties of the FIR/submm-selected galaxies in the local universe
may offer a valuable insight into interpreting the nature of these
highly obscured galaxies in the early universe.

In the study of an optically selected sample of 
299 normal and irregular galaxies, Condon, Anderson, \& Helou (1991a) 
found a {\it non-linear} relation between radio, FIR, and B-band 
luminosity, independent of Hubble type.  
Studies of an IRAS-selected galaxy sample by Lawrence et al. (1986;
303 galaxies in the North Galactic Pole), Unger et al. (1989; 156
galaxies in the South Galactic Pole), and Condon et al.
(1991a; 313 galaxies in the IRAS Bright Galaxy Sample [BGS]) extended
the observed tight correlation between radio and FIR luminosity to the
most luminous galaxies in the local universe.
The FIR emission of galaxies is sensitive to environmental effects,
and density inhomogeneity such as clustering can complicate the
derivation of a luminosity function (e.g. Oliver et al. 1996).  Therefore
a study involving the largest possible volume, bigger than large-scale
structures such as cosmic walls and voids, is highly desirable.  The
aim of this paper is to conduct a new extensive analysis of radio and
FIR emission in star-forming galaxies using the largest sample
available.  Our IRAS-selected sample with full redshift data is six
times larger (1809 galaxies) in size and five times deeper in distance
than all similar studies and covers over 60\% of
the sky.  Therefore our results should be more immune to such spatial
inhomogeneity and statistical fluctuations.  Both the normal late-type
galaxies and interaction/merger-induced high luminosity galaxies are
included in our sample, and possible differences between their
luminosity functions are examined.
 
An accurate derivation of the local radio and FIR luminosity functions also
allows the determination of the local radio and FIR {\it luminosity
densities} and thus an {\it extinction-free} estimate of the local
star formation rate (SFR).  For this reason, all cluster galaxies are
also explicitly included in our analysis, unlike some of the previous
studies.  Most existing estimates of the local SFR are based on either
UV continuum or optical emission line studies
\citep{Gallego95,Lilly96,Treyer98,Tresse98} that are subject to
highly uncertain extinction corrections
(e.g. Meurer et al. 1997, Steidel et al. 1999).  Optically thin FIR and radio
continuum emission provides an inherently {\it extinction-free}
estimate of star formation rate and a good calibration for the UV and
optical techniques.

The organization of the paper is as follows.  The sample selection and
the correlation of the radio and FIR properties are described in \S2.
The well known radio-FIR correlation is examined in detail, and
``radio-loud," ``radio-excess," as well as ``infrared-excess" galaxies
are identified in \S3 \& \S4.  The radio and FIR luminosity functions
for the local volume are derived, and inferences are made on the
nature of these luminosity functions in \S5.  Finally, the frequency
of the AGNs and their energetic importance are discussed, and the
infrared luminosity density is derived and discussed in terms of the
star formation rate in the local volume in \S6.  The strengths and
limitations of using the radio-continuum luminosity density for
measuring the star formation rate and cosmic star formation history
are also discussed.

\bigskip
   
\section{Sample Selection and Data Analysis}

The main catalog used for our sample selection is the 1.2 Jy IRAS
Redshift Survey catalog\footnote{The 1.2 Jy IRAS Redshift Survey
catalog was obtained from the Astronomical Data Center (ADC) operated
by NASA/Goddard Space Flight Center.}  (Strauss et al. 1990, 1992,
Fisher et al. 1995).  This catalog contains 9897 sources selected from the
IRAS Point Source Catalog (PSC) by the following criteria: $S_{60\mu
m}>1.2$ Jy, $(S_{60\mu m})^{2}>(S_{12\mu m}\times S_{25\mu m})$, and
$|b|>5^{\circ}$ \citep{Str92}.  Of the 9897 sources covering $87.6\%$
of the sky, approximately 5321 are galaxies and 14 were of unknown
type as of 1995 February 6.  Among the non-galaxian sources are HII
regions, stars, cirrus, and planetary nebulae.

The matching radio data for the FIR-selected sample were obtained from
the recently completed NRAO VLA Sky Survey \citep{Con98}.  The NRAO
VLA Sky Survey was made with D and DnC configurations of the Very
Large Array at a frequency of 1.4 GHz and an angular resolution of
$45''$.  Observations began in 1993 and extended into 1997, covering
the celestial sphere above $\delta=-40^{\circ}$.  The final NVSS
catalog contains $\approx 1.8\times10^{6}$ sources, including most of
the galaxies in the IRAS Faint Source Catalog, which has a flux limit
of $S_{60\mu m}\approx0.3$ Jy.

Both the IRAS redshift catalog and the NVSS catalog have well-known
limitations in their completeness (see above).  To construct an
unbiased complete sample for statistical analysis, we
selected objects at $\delta\ge-40^{\circ}$ and located
away from the Galactic plane ($|b| \geq 10^{\circ}$).  The redshift data
are complete only for the subsample with $S_{60\mu m}\ge2$ Jy
\citep{Str92}, and we adopted this flux cutoff for completeness.  The
sensitivity of the NVSS survey is about 0.5 mJy beam$^{-1}$
($1\sigma$), sufficient to detect all sample galaxies with
S/N $>20$.

We first constructed a catalog of FIR-selected galaxies with 1.4 GHz
radio counterparts by cross-correlating the 9897 sources in the 1.2 Jy
IRAS Redshift Survey with the radio sources in the NVSS database using
a program based on a software written by W. Cotton.  We sought radio
counterparts in the NVSS catalog within 30\arcsec\ of the IRAS PSC
positions, and unique counterparts were found in nearly all cases.
This search radius is twice the size of the typical 1$\sigma$ error
ellipse in the IRAS PSC, and the probability of finding an unrelated
NVSS source within each search radius is less than $6\times 10^{-4}$
(see Langston et al. 1990).  Two or more radio counterparts were found in
19 cases -- about three times larger than expected from chance
superposition.  Examination of these multiple matches revealed that
they are mostly multiple NVSS components produced by a single 
galaxy with a large angular extent (e.g., M82).
Because the NVSS is a snapshot survey, some of the extended disk emission
in galaxies with a large angular extent was missed.  Therefore the 1.4
GHz fluxes for the 134 large late-type galaxies and multiple matches
were replaced with the measurements made in the multiple-snapshot
study of bright spiral galaxies and the IRAS BGS galaxies by Condon
(1987) and Condon et al. (1990).  Only 25 cases required corrections
larger than 30\%.  The IRAS fluxes are from the PSC, but those for
the galaxies with optical diameter larger than 8$'$ are replaced
by the imaging measurements by Rice et al. (1988).

The 1.4 GHz and 60 $\mu$m luminosities were calculated using the 
following relations:
 $$log~L_{1.4 GHz} ({\rm W~Hz}^{-1}) = 20.08 + 2logD + log S_{1.4GHz}\eqno(1)$$
 $$log~L_{60\mu m} (L_\odot) = 6.014 + 2logD + log S_{60\mu m}\eqno(2)$$
where D is the distance in Mpc\footnote{$H_\circ = 75$ km s$^{-1}$ 
Mpc$^{-1}$ is assumed throughout this paper.}, and  
$S_{1.4 GHz}$ and $S_{60\mu m}$ are flux densities in units of Jy.  
This 60 $\mu$m luminosity ($L_{60\mu m}$) is the
luminosity contribution from the IRAS 60 $\mu$m band
to the FIR luminosity $L_{FIR}$\footnote{The IRAS 100 $\mu$m flux
is not available for a small subset of the IRAS redshift catalog sources, 
but all such sources are excluded from our complete sample by other
selection criteria.}, i.e. 
$$L_{FIR} (L_\odot)
=(1+{{S_{100\mu m}}\over{2.58 S_{60\mu m}}}) L_{60\mu m}\eqno(3)$$
(see Helou et al. 1988) and is related to another commonly cited quantity 
``$\nu L_{\nu}(60~\mu m)$" (e.g. Soifer et al. 1987, Saunders et al. 1990) as
$L_{60\mu m}= 0.69\times \nu L_{\nu}(60~\mu m)$.

Our final catalog of the IRAS 2 Jy Galaxy Sample sources meeting our
selection criteria contains 1809 galaxies.  The names, radial
velocities, optical magnitudes, 1.4 GHz and 60 $\mu$m luminosities,
and FIR/radio flux ratios $q$ (see Eq.~5) of the 161 luminous FIR
galaxies having $L_{60\mu} \ge 10^{11.3} L_\odot$ are given in
Table~\ref{tab:catalog}.  
We plan to publish the full catalog in an electronic form elsewhere.

The distribution of 1.4 GHz radio luminosity for the IRAS 2 Jy sample
is shown as a function of galaxy redshifts in Figure~\ref{fig:lzplot}.
Low-luminosity galaxies drop out quickly with increasing redshift due
to the IRAS 60 $\mu$m flux cutoff at 2 Jy, but the infrared luminous
galaxies are present to $z = 0.16$.  Therefore, our IRAS 2 Jy sample
is not only 6 times larger in the sample size compared with similar
previous studies, it also includes sources 5 times further away in 
the redshift domain as well.  
The {\it bona fide} radio-loud objects in the sample are PKS~1345+12,
3C~273, and NGC~1275, all of which stand out clearly in
Figure~\ref{fig:lzplot} with $L_{1.4 GHz} \geq 1.5\times10^{25}$ W
Hz$^{-1}$.
Determining the fraction of infrared-luminous
objects that are also radio loud (due to the presence of an 
AGN) is one of the key questions we address later (see \S4).
One immediate result from this study is the determination of accurate
positions for the IRAS Redshift Survey sources.  The positions of IRAS
sources such as found in the IRAS Point Source Catalog are only
accurate to about 15\arcsec\ (1$\sigma$).  In comparison, the source
coordinates given by the NVSS catalog have $\sim1''$ rms
uncertainties, sufficient for a secure identification of the
sources even in a crowded or confused field.  For example, the IRAS
position error ellipse for the interacting pair IRAS~12173$-$3514 falls
between the two widely separated optical galaxies (see
Figure~\ref{fig:i12173}) while the NVSS catalog identifies the
northern galaxy as the dominant source.

The histogram of position offsets between the IRAS PSC and NVSS 
(Figure~\ref{fig:offsets}) shows that the position offset is 
less than 10\arcsec\ in over 2/3 of all cases, and this is consistent 
with the expectation from the IRAS position error.   

\section{Radio-FIR Correlation}

The correlation between the radio and FIR luminosity in galaxies is
one of the tightest and best-studied in astrophysics.  It holds over
five orders of magnitude in luminosity \citep{Pri92}, and its nearly
linear nature is interpreted as a direct relationship between star
formation and cosmic-ray production \citep{harwit75,rickard84,deJ85,Hel85,Wun88}.

Evidence for non-linearity in this relation was discussed by Fitt, 
Alexander, \& Cox (1988) and Cox et al. (1988) for magnitude-limited 
samples of spiral galaxies.  Price \& Duric (1992) later showed that 
radio emission can be separated into free-free
and synchrotron emission and suggested that 
the latter causes a greater than unity slope at lower frequencies.  

\subsection{Observed Radio-FIR Correlation}

The plot of 1.4 GHz radio luminosity versus 60 $\mu$m FIR luminosity 
for our IRAS 2 Jy sample galaxies (Figure~\ref{fig:lums}a) shows a  
nearly linear correlation spanning over 5 decades in luminosity.
The three noteworthy features are: (1) a unity slope;
(2) steepening of the relation at low luminosity ($L_{60\mu m} \le
10^9 L_{\odot}$); and (3) a higher dispersion for luminosities
above $L_{60\mu m} \ge 10^{10.5} L_{\odot}$.  

A formal fit to the observed radio-FIR luminosity correlation yields
$$log(L_{1.4GHz})= [0.99\pm0.01] log(L_{60\mu m}/L_{\odot}) + [12.07\pm0.08].\eqno(4)$$
The overall trend is indistinguishable from a pure linear relation
since the overwhelming majority of FIR-selected galaxies have
$L_{60\mu m}$ between $10^{9-11.5} L_{\odot}$.  The observed strong
correlation is accentuated by the distance effect, but the plot of the
observed radio and 60 $\mu$m fluxes (Figure~\ref{fig:lums}b) also
shows a linear trend spanning nearly three orders of magnitude in
flux.  In both plots, the scatter is about 0.26 in dex.

There is a systematic tendency of galaxies with $L_{60\mu m} \le 10^9
L_{\odot}$ to appear below the best-fit line 
in Figure~\ref{fig:lums}a, and the diminished radio
emission of these galaxies may contribute to the previously reported
``non-linear" trend in the radio-FIR relation among optically
selected galaxy samples (e.g. Condon et al. 1991a).
Such a deviation from the linear relation can occur if the FIR or
radio luminosity is not directly proportional to the star forming
activity.  Dust heating by low mass stars (``cirrus" emission) may
contribute only to the observed FIR emission
\citep{Hel86,Lonsdale87,Fit88,Cox88,Dev89}.  Alternatively, the radio
luminosity of less luminous (and generally less massive) galaxies
may be low if cosmic-ray loss by diffusion is important
\citep{Klein84,Chi90}.  Because this sample is FIR selected, there
is a potential bias towards sources with higher FIR/radio flux
density ratios (see Condon \& Broderick 1986).  On the other hand,
low FIR luminosity sources with relatively high FIR/radio ratios may 
be missed systematically, possibly offsetting the first potential bias.

\subsection{Deviation from the Linear Relation}

The presence of any non-linearity or any
increase in the dispersion of the radio-FIR correlation
may be seen more easily  
by examining the ``$q$" parameter (Condon et al. 1991a),
  $$q \equiv log[{{FIR}\over{3.75 \times 10^{12} {\rm W~m}^{-2}}}] - log[{{S_{1.4 GHz}}\over{{\rm W~m}^{-2} {\rm Hz}^{-1}}}]\eqno(5)$$
where $S_{1.4 GHz}$ is the observed 1.4 GHz flux density in units of 
W m$^{-2}$ Hz$^{-1}$ and
$$FIR \equiv 1.26 \times 10^{-14}(2.58S_{60\mu m}+S_{100\mu m})~{\rm W~m}^{-2} ,\eqno(6)$$
where $S_{60\mu m}$ and $S_{100\mu m}$ are IRAS 60 $\mu$m and 100
$\mu$m band flux density in Jy (see Helou et al. 1988).  Therefore $q$ is
a measure of the logarithmic FIR/radio flux-density ratio.  For most
galaxies in the IRAS Bright Galaxy Sample $q \approx 2.35$, but
some galaxies have smaller $q$ values due to an additional
contribution from compact radio cores and radio jets/lobes
\citep{San96}.  Optically selected starburst galaxies
(interacting/Markarian galaxies) have about the same $q$ values as normal
galaxies, and the radio-FIR flux ratio appears to be independent of
starburst strength (e.g. Lisenfeld, Volk, \& Xu 1996).

The $q$ values for the IRAS 2 Jy sample were computed using Eq.~5 and are
shown in Figure~\ref{fig:qplot}.  The mean $q$ value for the entire
sample is $2.34\pm0.01$, 
which is in good agreement with $q$
values typically found in other star-forming galaxies.  In fact,
the 98\% of galaxies in our IRAS 2 Jy sample are found between
the two dotted lines which mark the five times excess and deficit 
of radio emission with respect to mean $q=2.34$ (solid horizontal line).

The effects of measurement errors and source confusion can be 
estimated from the dispersion in the $q$ value in Figure~\ref{fig:qplot}.  
Before the radio fluxes for galaxies with large 
angular extent were corrected using the measurements by Condon (1987),
several objects with unusually large $q$ values ($q\ge3$) were
found -- resulting from the missing extended flux.  
One object, NGC~5195, also turned out to 
include incorrect IRAS fluxes in the IRAS redshift catalog, confused
with 
its brighter companion M51.  Using the correct IRAS fluxes, its 
$q=2.68$ became much closer to the sample mean value.  
Similarly, IRAS~02483+4302 is a well
known quasar-galaxy pair and another case of source confusion.
Its 1.4 GHz flux was corrected using the measurement by
Crawford et al. (1996).  These occurrences are consistent 
with the expected source confusion statistics discussed earlier (\S2.3).  
The direct comparison of Figure~\ref{fig:qplot} with a similar
plot in Condon et al. (1991a) suggests that the measurement errors
and confusion effects are nevertheless small and comparable 
to those of the bright spiral sample
and the IRAS BGS sample analyzed by Condon et al.

There are good reasons to believe that galaxies with much larger or
smaller $q$ values are inherently different.  Firstly, there are
several objects whose smaller $q$ values are the
result of significant excess radio emission associated with AGNs (see
\S4 for a further discussion).  There are also a few infrared-excess
objects which may be highly obscured compact starbursts or dust
enshrouded AGNs (see \S6.1 \& \S6.2), and some of the observed dispersion may
arise at least in part from the variation in excitation or dispersion
in dust temperature.  The clustering of low-luminosity galaxies
($L_{60\mu m} \simless 10^9~L_\odot$) above the mean $q=2.34$ in
Figure~\ref{fig:qplot} is another clue that the scatter in the $q$
distribution is more than simply statistical in nature.  
This stems from the previously noted deviation from the linear
radio-FIR relation (\S3.1), and it is also the source of a small
gradient in $q$ noted by Condon et al., due to a systematic decrease
in radio emission among low luminosity galaxies.  Lastly, the
dispersion in the $q$ values among the objects with $L_{60\mu m}
>10^{11} L_\odot$ is significantly larger, $\sigma_q=0.33$.  This
larger scatter also manifests itself in Figure~\ref{fig:qplot} as the
disappearance of the dense core in the distribution near the mean.
Helou et al. (1985) noted a similar increase in dispersion at high
luminosity in their analysis of only 38 spiral galaxies and suggested
that the radio-FIR correlation may break down for high luminosity
sources.  The new data suggest that the correlation still holds but
the dispersion is indeed increased (see below for possible explanations).

\bigskip

\section{Radio-Loud and Radio-Excess Galaxies}

Two different functional definitions of ``radio-loud" objects are
found in the literature: one based on absolute radio power and 
the second based on a relative flux ratio (e.g. $L_{radio}/L_{opt}$).  
The former are generally a subset of the latter.  Here we call
``radio-loud" only those galaxies hosting
radio sources with $L_{1.4 GHz} \ge10^{25}$ W Hz$^{-1}$.  
In contrast, we define ``radio-excess" objects as galaxies
whose radio luminosity is five times the value predicted by the
radio-FIR correlation or larger (below the lower dotted line in 
Fig.~\ref{fig:qplot}, i.e. $q\le 1.64$).  

\subsection{Radio-Loud Objects}

It is clear from Figure~\ref{fig:lzplot} that there are only three 
radio-loud objects in our sample of 1809 galaxies, with 
several others approaching the radio-loud limit.  
{\it Regardless of the exact statistics, the {\it bona fide} radio-loud
galaxies are extremely rare among the infrared selected sample}.
We discuss below the three radio-loud objects in our original sample in 
detail. It appears that each object represents a special case, and
their radio and FIR emission have physically distinct origins.

\noindent{\bf PKS~1226+023 (3C~273)} is a well-known 
QSO at $z=0.158$ (Schmidt 1963).  
It is a warm IRAS source ($S_{60\mu m}/S_{100\mu m}=0.76$, 
$T_{dust}\sim 44$ K) and also the most luminous 
(log $L_{1.4GHz}=27.42$ W Hz$^{-1}$) radio source in our sample.  
Its elliptical host is somewhat brighter than the 
brightest galaxy in a rich cluster ($M_V= -22.1$, Bahcall et al. 1997) 
and is probably located inside a poor cluster (Stockton 1980).  
The large-scale radio structure from the VLA and MERLIN shows a compact, 
flat-spectrum core and a single jet extending about 23$''$ 
from the core at a position angle of 222$^\circ$ (Conway et al. 1993).
Unlike the other two radio-loud objects discussed below, its infrared
emission appears to be mostly the continuation of the non-thermal
spectrum from optical to radio wavelengths, associated directly with
the AGN.

\noindent{\bf NGC~1275 (3C~84)} 
is the central giant elliptical galaxy of the Perseus
Cluster with a Seyfert~2 nucleus and may be accreting gas
from the X-ray emitting intercluster medium via a cooling flow 
\citep{Kent79,Fabian81}.  The bulk of its 1.4 GHz radio continuum 
emission (log $L_{1.4GHz}=25.13$ W Hz$^{-1}$)
comes from the core component less than 1\arcsec\ in size,
but it also has a 10\arcmin\ size extended radio structure 
that may be interpreted as an asymmetrical Fanaroff-Riley type I
source whose jet axis lies close to the line-of-sight 
\citep{Miley75,Ped90}.  
The FIR emission is attributed to an extended ($\sim30''$) 
distribution of dust, possibly from a kpc scale circumnuclear 
gas/dust structure accreted from a cluster interloper 
\citep{Les95,Inoue96}.  Unlike the compact nuclear radio source, 
the 100 $\mu$m emission in NGC~1275
shows little evidence for variability \citep{Les95}.
Therefore, the FIR emission in NGC~1275 may be
unrelated to the radio emission, which is dominated by the 
non-thermal emission from the AGN.

\noindent{\bf PKS~1345+12 (4C~12.50)} is the second
most luminous radio source in our sample 
(log $L_{1.4GHz}=26.18$).  The optical galaxy
has a double nucleus embedded in a distorted halo
$\sim85$ kpc in diameter, indicating a merger system 
approaching the appearance of a cD galaxy \citep{Mirabel89,Sha92}.
This galaxy has a core-halo radio structure, and more than 95\% of 
the flux at 2 cm wavelength is coming from a region smaller than 
$0''.05$ (106 pc; van Breugel et al. 1984).
The western nucleus shows a Seyfert 2 spectrum \citep{Gil86},
and Veilleux et al. (1997) reported broad infrared recombination 
lines indicative of a hidden QSO.  
The eastern nucleus is associated with a compact 
steep-spectrum radio source with a double-lobe structure 
(Shaw et al. 1992).  The large infrared luminosity 
(log $L_{FIR}(L_\odot)=11.9$)
makes it a genuine ultraluminous infrared galaxy with
an extremely large gas content inferred from the CO observations
($M_{H_2}=6.5\times 10^{10}~M_\odot$; Mirabel et al. 1989, Evans
et al 1999). Therefore it is one of the best candidates for studying 
a possible common origin for ultraluminous galaxies and 
powerful radio galaxies.

\subsection{Radio-Excess Objects}

One of the key objectives of this study is a quantitative analysis of
the frequency and energetic importance of AGNs in
FIR-selected galaxies.  Assuming that radio excess is an
AGN indicator, a total of 23 potential AGN hosts with $q\le1.64$ 
are identified in our sample (see Table~\ref{tab:radio-excess}). The low
frequency (23/1809 = 1.3\%) of these radio-excess objects indicates
that radio AGNs occur quite rarely among the infrared selected 
sample of normal and starburst galaxies.  
Among the FIR-luminous galaxies with log~$L_{FIR}\ge 11.3 L_\odot$ 
listed in Table~\ref{tab:catalog}, there
are only three radio-excess galaxies (IRAS~12173$-$3541,
IRAS~12265+0219, IRAS~13451+1232).  This fraction, 3/161 = 1.9\%, is
only marginally larger than the radio-excess fraction among the entire
IRAS selected sample.  {\it Therefore the radio-excess fraction is
generally only a few percent and is nearly independent of the
infrared luminosity.}

How does this result compare with the analysis of optically selected
galaxy samples?  Condon \& Broderick (1988) have identified a total of
176 radio sources with $S_{1.4 GHz}\ge 150$ mJy among galaxies in {\it
Uppsala General Catalogue of Galaxies} (Nilson 1973; hereafter UGC).
Judging by the radio morphology, radio-to-infrared flux density ratio,
and infrared spectral index, 64 are identified as ``starbursts" and
103 (59\%) are identified as ``monsters".  There are four borderline
cases while no infrared data are available for the remaining three.
We find that the same distinctions could have been made using their
radio-to-FIR flux ratio alone.  Among the ``monsters" in the UGC
sample, only 6 galaxies (UGC~2188, 2669, 3374, 3426, 8850, and 12608)
are bright enough at 60 $\mu$m to be included in our sample, and all
but UGC~2188 are identified as radio-excess by our analysis.  The
starburst/Seyfert galaxy UGC~2188 (NGC~1068) has $q=1.72$ and is also
marginally radio-excess (about 4 times larger radio luminosity than
expected from the radio-FIR correlation).  Therefore there is little
overlap between the ``monsters" in the UGC sample and our IRAS 2 Jy
sample.  This is mainly due to the relatively large radio flux cutoff
used in the UGC study, which has introduced a strong bias in favor of
radio bright AGNs.  As the radio flux cutoff is lowered, the dominance
of the star-forming galaxies emerges quickly as shown by the
preliminary analysis of the UGC galaxies using the NVSS database
\citep{cotton98}.  The relative frequencies of ``monsters'' and
star-forming galaxies is addressed further in
terms of their luminosity functions below.

The frequency of radio-excess objects derived above is
strictly a lower limit to the overall frequency of AGNs among the 
IR selected galaxies.  Radio luminosity of AGNs ranges over 
10 orders of magnitudes (e.g. Condon \& Broderick 1988, 
Ho \& Ulvestad 2000), and radio-quiet AGNs might not be that rare 
in our sample.  Nevertheless a large majority of luminous
AGNs are radio sources and may be readily identified. 
For example, 84\% (96/114) of Palomar Bright Quasar Survey 
sources have been detected by Kellermann et al. (1989)
at 5 GHz using the VLA.  Evidence for radio emission in excess of
the inferred star formation activity level among 
high redshift dusty, radio-quiet QSOs has also been
discussed by Yun et al. (2000).  To quantify the robustness of
radio-excess as an AGN indicator, we have examined 
the radio properties of a sample of 48 Palomar-Green (PG) 
survey QSOs with IRAS detections 
reported by Sanders et al. (1989).  Ten of these sources 
were dropped from our analysis, either because the XSCANPI analysis
revealed the IRAS detections to be the results of confusion or
because meaningful upper limits for the FIR and radio luminosity
could not be obtained.
Among the remaining 38 PG QSOs, only 23 have $L_{60 \mu m}>10^{11} 
L_\odot$.  The radio-excess fraction among this subsample of
FIR luminous QSOs is 12/23=52\%, which greatly exceeds the 
frequency for our IRAS selected galaxy sample 
(see above).\footnote{The radio-excess fraction for the entire
range of $L_{60 \mu m}$ is between 34\% and 45\%, depending on 
how the sources with only upper limits of $q$ are accounted.}
Among the remaining 11 PG QSOs without the
excess radio emission, Alloin et al. (1992) detected CO emission 
with inferred molecular gas mass larger than $10^{10} M_\odot$ in
all three objects observed and argued that star formation
may be responsible for the bulk of the FIR luminosity among these
QSOs.  
Therefore the overall frequency of AGNs may be somewhat larger
than inferred from the radio-excess alone --  perhaps as much
as twice as large in the case of the IRAS selected PG QSOs. 
Further discussions of the frequency of energetically dominant 
AGNs are found in the Discussion section (\S6), including the 
examination of the mid-infrared properties.
\bigskip

\section{Radio and Far-infrared Luminosity Functions}

With 1809 galaxies in our IRAS 60 $\mu$m flux-limited sample, we
should be able to derive the radio and FIR luminosity functions (LFs)
for field galaxies in the local volume with significantly better
statistics than previously possible.  Since our sample covers a
significantly larger solid angle and volume than the samples used
previously, fluctuations produced by large-scale structures and
clustering should be smaller.  
If there is a significant presence of radio AGNs contributing to our
IRAS selected sample, not only they can be identified as a separate
population in the derived radio and/or FIR luminosity functions, we
can also determine the range of luminosity where their contribution is
significant.

Several different functional forms have been suggested previously 
for the radio and FIR luminosity functions of field galaxies, such as  
log-normal \citep{Hum81,Iso92}, hyperbolic \citep{Con84}, 
two power-laws \citep{Soifer87}, 
and power law/Gaussian \citep{San79,Sau90}.  These functional 
forms offer reasonable descriptions of the observed radio and FIR 
LFs with varying degrees of relative merit.   
All previously published studies, however, have unanimously  
rejected the traditional form proposed by Schechter (1976) as being too 
narrow to describe the observed radio or FIR LFs.  This is  
surprising because Schechter function offers an excellent 
and robust description of the optical LFs of field galaxies and 
galaxy clusters (e.g. Shapiro 1971, Schechter 1976, 
Auriemma et al. 1977, Efstathiou, Ellis, \& Peterson 1988).  
Extinction effects at optical wavelengths cannot explain this
discrepancy because Schechter function also
offers an excellent description of near-IR LFs of galaxies  
\citep{Gar97,Szo98,Love00}.  Therefore the failure of the Schechter
form is highly unusual and deserves further investigation.

\subsection{Schechter Luminosity Function \label{sec:Schechter}}

Here we revisit the idea of describing the observed radio and FIR 
LFs in terms of Schechter functions for two important reasons.  
First, the Schechter function has a simple but sound theoretical
basis as it follows from a theoretical analysis of self-similar 
gravitational condensation in the early universe \citep{Press74}.
The Schechter function has a form
$$\rho(L)dL = \rho^{*}(L/L^{*})^{\alpha}\exp(-L/L^{*})d(L/L^{*})\eqno(7)$$
where $\rho^{*}$ and $L^{*}$ are the characteristic density and 
luminosity of the population, and $\alpha$ describes the 
faint-end power-law slope for $L\ll L^{*}$ (Schechter 1976).  
Not only is the Press-Schechter formalism a well accepted paradigm 
for cosmological structure formation, but the Schechter function also
describes the observed optical and near-IR LFs of field and 
cluster galaxies well as already mentioned.  
If star formation activity is an integral part of galaxy 
evolution and reflects the nature of the host galaxies, then  
the Schechter function should also offer 
a good description for the radio and FIR LFs as it does
for the optical and near-IR LFs.

Another reason for reconsidering the Schechter form is that the  
observed departure, namely the power-law tail at high luminosity end
for the FIR LF, may have been a {\it predictable consequence} of the 
underlying theoretical consideration in the Press-Schechter formalism.
To be specific, the Press-Schechter formalism for CDM 
structure formation predicts a {\it mass function} of the Schechter
form, rather than a luminosity function.  To make the logical connection 
between the
observed LFs and the Press-Schechter formalism, one has to invoke 
a constant mass-to-light (M/L) ratio, which is a reasonable
assumption at the optical and near infrared wavelengths. 
For starburst galaxies, on the other hand, a momentary jump in 
L/M by 2-3 orders of magnitudes is expected by
definition, nearly entirely at the FIR wavelengths.  Therefore,
if the underlying mass distribution is of a Schechter form,
the resulting FIR (and radio) LF should be a linear combination of
a Schechter function (for field galaxies) and a high luminosity excess
due to starburst galaxies.  Hence, we pose a hypothesis that 
the observed radio and FIR LFs are indeed linear sums of two 
Schechter functions, one for the field galaxies and second 
for the starburst population.  Choosing a Schechter form for
the starburst population is largely for self-consistency.
If this population is distinguished only by 2-3 orders of
magnitude elevation in M/L, then a Schechter function should also  
represent a reasonable form for its LF.  We examine this hypothesis by 
analyzing the derived radio and FIR LFs in detail below.

The characteristic Schechter parameters for the observed LFs
are derived in two different ways.  A luminosity function is 
the {\it space density} of the sample galaxies within a
luminosity interval of $\Delta L$ centered on $L$. 
Thus the LF and associated uncertainty can then be derived ``directly" as
$$\rho_{m}(L) = \sum_{i=1}^N ({{1}\over{V_m}}),~~~
\sigma_{\rho(L)} = (\sum_{i=1}^N {{1}\over{V_m^2}} )^{1/2}\eqno(8)$$
where $\rho_{m}(L)$ is the measured luminosity density within
the ``magnitude" bin centered on luminosity $L$ and $V_m$ is
the sample volume appropriate for each of the N=1809 galaxies
\citep{Sch68,felten76}.  
The sample volume $V_m$ was determined by the IRAS 60 $\mu$m sensitivity 
limit of 2 Jy and the NVSS survey area as:
$$\int_{0}^{D}r^{2}dr\int_{0}^{2\pi}d\rho
   \int_{\delta=-40^\circ}^{\delta=+90\circ} 
   \cos\delta d\delta = 3.44D^{3}.\eqno(9)$$
The sampled volume is further reduced by 18\% due to the exclusion
of the Galactic Plane and the omitted areas in the IRAS redshift survey.

The Schechter parameters can also be derived from the observed
luminosity distribution, $n(L)$.
For a randomly chosen sample, the distribution function $n(L)$ is given by
     $$n(L) \equiv \rho(L)V(L),\eqno(10)$$
where $V(L)$ is the volume sampled.  The distribution function 
can then be re-written as
$$n(L)dL =\rho^{*}V^{*}(L/L^{*})^{\alpha+3/2}
           exp(-L/L^{*})d(L/L^{*})\eqno(11)$$
where $V^{*}$ is the characteristic volume associated with $L^{*}$.
Given these analytical expressions for the luminosity distribution
and luminosity function, a $\chi^{2}$ 
analysis is performed on the observed $\rho(L)$ and
$n(L)$ to deduce the characteristic Schechter parameters.  
The Schechter parameters derived from 
$\rho(L)$ and $n(L)$ are slightly different because the two 
methods are subject to slightly different systematic effects.  In addition
to the formal uncertainties given by the $\chi^{2}$ analysis, the 
difference between the best-fit parameters derived by the two 
different methods offers a quantitative measure of the
systematic errors in the derivations.

A potential source of confusion in comparing the
derived LFs from one study to another is the definition of luminosity 
interval $\Delta L$ used.  Both ``per magnitude" and 
``per dex" are commonly used in the literature.  We
adopt ``per magnitude (mag$^{-1}$)" in our derivations of 
radio and FIR luminosity functions.  All published luminosity
functions using ``dex$^{-1}$" intervals (e.g. Saunders et al. 1990)
are thus corrected by a factor of 2.5 before making any comparisons.

\bigskip

\subsection{IRAS 60 $\mu$m Luminosity Function}

The IRAS 60 $\mu$m luminosity distribution and luminosity function 
for the IRAS 2 Jy sample galaxies are constructed using 18 
luminosity bins of log $L_{60 \mu m}$ ($L_\odot$) between 7.75 and 12.25
with a width of $\Delta \log L = 0.25$. 
The computed 60 $\mu$m luminosity distribution and LF are tabulated in
Table~\ref{tab:60LF} 
along with the computed $V/V_m$, which is a measure of the
sampling completeness \citep{Sch68,felten76}.  These quantities are
also shown graphically in Figure~\ref{fig:60LF} along with the
best-fit Schechter functions (see below).  The luminosity distribution
shown in Figure~\ref{fig:60LF}a nicely indicates that the most
common objects in the IRAS 2 Jy sample have 60 $\mu$m luminosity of
about $10^{10} L_\odot$, typical of $L^*$ galaxies in the field.

The derived 60 $\mu$m LF is shown in Figure~\ref{fig:60LF}b
and is compared with previous derivations by Soifer et al. (1987)
and Saunders et al. (1990).  The general shape of the LF
may be reasonably described as a power-law distribution, but two or
more power-laws are needed to account for the detailed shape
in the entire luminosity range.  As discussed in the
previous section (\S~5.1), a linear sum of two Schechter functions may 
make a good physical sense as a description for the radio and
FIR luminosity functions, and indeed the derived analytical
expressions shown in the solid and dashed lines match the 
the observed 60 $\mu$m LF closely.  In deriving the best fit Schechter
parameters, the faint end power-law slope $\alpha$ is assumed to
be universal because it cannot be derived cleanly for the
higher luminosity (``starburst'') population.
The characteristic luminosities $L^*$ derived from the 60 $\mu$m
luminosity distribution (Figure~\ref{fig:60LF}a) for the high- and
low-luminosity populations are $(21\pm2)\times 10^{10} L_\odot$
and $(2.2\pm0.4)\times 10^{10} L_\odot$, respectively, with the
faint end power-law slope $\alpha= -0.81\pm0.06$.  The 
best fit parameters can also be derived from the observed
60 $\mu$m luminosity functions (Figure~\ref{fig:60LF}b), and they are
$(22\pm2)\times 10^{10} L_\odot$ and $(2.3\pm0.5)\times 
10^{10} L_\odot$, with $\alpha= -0.82\pm0.05$ 
(see Table~\ref{tab:60LF.fits} for the summary and the ${\chi}^2$
analysis).  

A break in the FIR LF has previously been suggested by 
Smith et al. (1987) and Saunders et al. (1990) based on the 
optical morphology and infrared color.  These authors had not offered
any clear physical explanations for such a break seen near
$L_{60\mu m}=10^{11} L_\odot$, however.  The combination of an even 
cleaner break seen in the 1.4 GHz radio LF
(see below) and the excellent fits reported in 
Table~\ref{tab:60LF.fits} lend strong support for the two
Schechter function description 
and the implied physical explanation in terms of an elevated 
mass-to-light ratio for the starburst population.

Significant differences are seen among the several 60 $\mu$m
LFs compared in Figure~\ref{fig:60LF}b at the low luminosity end.
Our 60 $\mu$m LF (filled circles) is smoother and has smaller 
error bars than those derived by Soifer et al. (1987) and by
Saunders et al. (1990) because our sample size is 6 times larger.   
Our 60 $\mu$m LF dips below the Soifer et al. LF (empty squares) 
and the derived Schechter model near $L_{60\mu m}\simless 10^{8.5} L_\odot$.
The 60 $\mu$m LF derived by Saunders et al. (1990; empty circles)
flattens even further, beginning around log $L_{60\mu m} \sim 9.5$.  
This flattening is also seen in their Figure~2, in comparison with
that of BGS sample (Soifer et al. 1987), Smith 2 Jy sample 
(Smith et al. 1987), and Strauss \& Huchra (1988).  
One explanation for the different faint end slope may be the
sample completeness as the surveys with particular emphasis
on the completeness (e.g. Smith et al., Stauss \& Huchra) 
tend to derive a steeper faint end slope.  In contrast,
Saunders et al. data probably suffers the worse completeness
problem because their luminosity function is derived
using a heterogeneous collection of several different surveys.
The plot of $V/V_m$ (Fig.~\ref{fig:60LF}c) shows that our sample
is distributed fairly uniformly within the sampling volume
($V/V_m \sim 0.5$).  

The derived $V/V_m$ for our sample dips below
0.5 at log $L_{60\mu m} \simless 8.5$, albeit with large errorbars.
This can be interpreted as an indication of local density 
enhancement and may be reflecting our special location 
within the local galaxy group and the local super-cluster. 
Derivation of LFs using the $1/V_m$ method (Eq.~8)
has been suggested to be susceptible to
the large scale inhomogeneity, and the steep rise in the
derived LF at the faint end may be an artifact of this bias.
The steepest rise in the faint end slope is associated with
the LF derived by Soifer et al. which also has the smallest
sampling volume, in accordance with this expectation.  
Assuming the luminosity function has a universal form,
Yahil et al. (1991) proposed a density-independent determination
of the luminosity function by introducing a selection function
which accounts for the distance effect.  The Yahil et al.
derivation of a double power-law form LF using an earlier version of
the IRAS redshift catalog is shown as a dotted curve
in Figure~\ref{fig:60LF}b.  While the derived
faint end slope is flatter and shallower than the LF we derive,
it is also steeper and higher than the LF derived by Saunders et al. 
(also see their Fig.~7).  Even such a density-independent 
method faces a fundamental limitation if the faint end slope is 
constrained by the low luminosity objects that are drawn 
mainly from the local volume with enhanced density.
Therefore the analysis presented here
suggests that the faint end behavior of the luminosity function
is intrinsically poorly constrained.
It should be noted, however, that these differences between the various
LFs at low L do not have much effect on the implied
star formation density discussed below (\S6.4)
since the star-formation rate is proportional
to the LF multiplied by L, a function peaking at higher luminosity.

\subsection{1.4 GHz Radio Luminosity Function}

Because our sample is selected by the IRAS 60 $\mu$m band flux
density, deriving the 1.4 GHz radio luminosity function 
for the same sample is not as straight forward as in the 60 $\mu$m
LF case.  In particular, the 1.4 GHz radio LF cannot be derived
from the luminosity distribution (Eq.~11) because the volume sampled
does not relate directly to the 1.4 GHz luminosity.  Instead,
the 1.4 GHz radio luminosity function is derived using the
sampling volume corresponding to the IRAS 60 $\mu$m luminosity
of each sample galaxy.  The logic behind this
calculation is that each sample galaxy contributes to the
local number density at their respective luminosity bins in
exactly the same manner for both the 60 $\mu$m and 1.4 GHz LF,
and sample selection is reflected identically by the
sample volume determined by the IRAS 60 $\mu$m flux limit.

The computed radio luminosity distribution and LF are tabulated in
Table~\ref{tab:1.4LF} along with the computed $V/V_m$.  
These quantities are also shown graphically in Figure~\ref{fig:1.4LF} 
along with the best-fit Schechter functions (see Table~\ref{tab:1.4LF.fits}).  
The radio luminosity distribution, LF, and the $V/V_m$ plots are
qualitatively similar to those of the IRAS 60 $\mu$m functions
in Figure~\ref{fig:60LF}, mainly because of the linear radio-FIR
correlation (see \S3 \& \S6.3).  
The derived 1.4 GHz radio LF, shown in Figure~\ref{fig:1.4LF}b,
is flat at low luminosity end
and falls off steeply beyond $10^{22}$ W Hz$^{-1}$.
A Schechter function (solid line) appears to be a good description for all
low luminosity objects of $L_{1.4 GHz}\le 10^{23}$ W Hz$^{-1}$, 
with the best fit characteristic Schechter parameters of 
$L^*=(2.1\pm0.3)\times 10^{22}$ W Hz$^{-1}$,
$\rho^*=(3.2\pm0.2)\times 10^{-4}$ Mpc$^{-3}$ mag$^{-1}$, 
and $\alpha=-0.63\pm0.05$.  
The observed radio LF is significantly flatter than the 
best fit Schechter function below $L_{1.4 GHz}\le 10^{20.4}$ W Hz$^{-1}$.
In addition to the possible effects of sample completeness and
local density enhancement, the systematic deviation from the
linear radio-FIR relation may further contribute to this flattening.

The presence of the luminous second population is more obvious in
Figure~\ref{fig:1.4LF}b at $L_{1.4 GHz}\ge 10^{23}$ W Hz$^{-1}$.  A
full $\chi^2$ analysis is not warranted for this population because of
the small numbers of bins represented and the large observed scatter.
Characteristic luminosity and density of
$L^*=(1.4\pm0.2)\times 10^{23}$ W Hz$^{-1}$ and
$\rho^*=(8.3\pm0.8)\times 10^{-6}$ Mpc$^{-3}$ mag$^{-1}$ are derived
assuming the same faint-end power law slope of $\alpha=-0.63$,
but these numbers should be taken only as illustrative
values.  The flux cutoff of 2 Jy at 60 $\mu$m and
the resulting volume sampled are probably not sufficient to fairly
represent these rarest objects with high luminosity.

The comparison of our radio LF (filled circles) 
with that of the flux limited
sample of UGC galaxies by Condon (1989; open circles) 
shows that the low radio luminosity galaxies in our IRAS 
selected sample are nearly identical to the ``starburst" galaxies
(late type galaxies in the field, as opposed to ``monsters")
in the UGC catalog (see Figure~\ref{fig:1.4LF}b).  The small
excess associated with the IRAS 2 Jy sample is entirely accounted
by the cluster galaxy population.  When all galaxies within the 
3 Mpc projected diameter of the known clusters within 100 Mpc 
distance are removed, the resulting radio LF is indistinguishable 
from the UGC sample LF.  The significance of this result is that
{\it there is a complete overlap between the optically selected
late type galaxies and IRAS selected galaxies in the field}.
In other words, the measured radio luminosity may be used 
to infer star-forming activity for late-type galaxies in the
field with the same level of confidence as using their FIR 
luminosity.  Estimating the star-formation rate for an individual
galaxy using its radio luminosity may still be uncertain 
by the magnitude of the observed scatter in Figure~\ref{fig:qplot},
but such an analysis for an ensemble of galaxies should be 
significantly more reliable (see \S6.5).

The comparison with the ``monsters" in the UGC sample (open squares), 
which has a flat luminosity function, is more complicated.
The 1.4 GHz radio LF for the ``monsters" derived 
by Condon is comparable to that of our IRAS selected galaxy sample 
for $L_{1.4 GHz}$ between $10^{23}$ and $10^{24}$ W Hz$^{-1}$, but
the difference quickly grows to more than an order of magnitude at 
$L_{1.4 GHz}\simgreat 10^{24}$ W Hz$^{-1}$.  This disagreement
at the highest luminosity is highly statistically significant, 
and it must be explained by the incompleteness
in the IRAS 2 Jy sample.  The LF of galaxies is a property
of galaxies and not of the sample; the $1/V_m$ weighting is
supposed to correct for sample selection effects.
With a sufficiently large sample of IRAS galaxies, the 
LF should equal the LF of optical/radio selected UGC galaxies,
including both starbursts and monsters.  The main reason
that our IRAS total LF is lower than the UGC monster LF
must be that NO ``monsters" have been detected by IRAS in the
highest luminosity bins.  If even one or two were detected, 
they would have small $V_m$ values and hence large weights,
bringing up the IRAS LF to agree with the UGC LF.
Therefore, the second, more luminous population in the IRAS 2 Jy
sample cannot be simply identified with the ``monsters" in the 
UGC sample.  Because these high luminosity objects also largely
follow the same radio-FIR correlation as the low luminosity objects
(see Fig.~\ref{fig:lums}), the observation can be explained
more naturally with our hypothesis 
that the high-luminosity excess represents a physically distinct 
second population, whose large luminosity is the manifestation
of a burst of tidally or merger induced activities (see 
\S~\ref{sec:Schechter}).  Possible contribution by luminous 
AGNs are discussed in greater detail below.

\bigskip

\section{Discussion \label{sec:discussion}}

\subsection{High Luminosity Objects and Frequency of AGNs}

Infrared-luminous galaxies with $L_{FIR}\ge 10^{11}~L_\odot$ 
(equivalently, $L_{1.4GHz}\ge 10^{23}$ W Hz$^{-1}$) are
special objects because few normal disk galaxies in the local 
universe have such large luminosity.  A 
complete list of 161 luminous infrared galaxies with 
$L_{60\mu}\ge 10^{11.3} L_\odot$ is given in Table~\ref{tab:catalog}.
In addition to many well-known ultraluminous infrared
galaxies (e.g. Arp~220), Seyferts, 
starburst, and HII galaxies are included in the list, and  
morphological descriptions of ``interacting", ``galaxy pairs", 
or ``mergers" are common.
More importantly, they may represent a distinct separate 
population from the late type field galaxies
in the derived radio and FIR LFs (see \S5).  

The infrared color-color diagrams shown in Figure~\ref{fig:color}
provide a
useful tool for probing the mean radiation field and 
dust composition.  In analyzing the
IRAS colors of field galaxies, Helou (1986) found a linear
trend of increasing dust temperature with increasing luminosity
as shown by the dashed parallel tracks in Figure~\ref{fig:color}.  
The high-luminosity 
objects (filled large circles) as a group show a systematically larger
60 $\mu$m to 100 $\mu$m flux ratio, which is an indication of warmer 
mean dust temperature
($S_{60\mu m}/S_{100\mu m}\sim 0.8$, $T_{dust}\sim 45$~K). 
They also cluster near the upper left corner of Figure~\ref{fig:color}, 
which indicates that the dust in these galaxies is
exposed to hundreds of times stronger radiation field
than that of the solar neighborhood.  This clustering is broadly
consistent with the intense starburst interpretation for these
FIR-luminous systems.  Even though their FIR luminosity is on
average 10 times smaller, the radio-excess objects (radio AGNs, shown 
as diamonds) show a similar IRAS color distribution as 
infrared luminous galaxies.  An AGN produces its characteristic mid-IR
enhancement by heating a small amount of dust in its immediate surroundings
to a very high temperature.  The observed similarity in the IR color between
luminous starbursts and AGNs might indicate that the warm dust heated
by stars may also be concentrated in a relatively small volume.

While an unambiguous distinction is not possible, a systematic
separation among the late type field galaxies, infrared luminous
galaxies, and radio AGNs becomes more apparent in the far-IR versus mid-IR
color plot as shown in Figure~\ref{fig:color2}.  The far-IR color of
the FIR-luminous sample is clearly and systematically warmer than
that of the late type field galaxies (small dots).  The median
$S_{60\mu m}/S_{100\mu m}$ ratio for the FIR-luminous galaxies is
about 0.8 (45 K for $\beta=1$; see Helou et al. 1988) while it is about
0.5 (36 K for $\beta=1$) for the field galaxies.  The radio-excess
objects show a similar range of the far-IR colors as the infrared
luminous objects.  In addition, they also show a systematic offset
towards the warmer mid-IR color by about 0.5 in dex, separating themselves 
from the infrared luminous galaxies.
Presence of hot dust in the circumnuclear region surrounding an AGN
offers a plausible explanation for this mid-IR enhancement.  Mid-IR
excess with $S_{25\mu m}/S_{60\mu m}\ge 0.18$ (right of the thin vertical
line) has been empirically proposed as an indicator of an
infrared AGN by de Grijp et al. (1985).  Indeed the radio-excess
objects in our sample show the largest $S_{25\mu m}/S_{60\mu m}$
ratios among all IRAS selected galaxies (see Figure~\ref{fig:color2}).

A total of 20 out of the 161 FIR-luminous objects in our
IRAS 2 Jy sample have $S_{25\mu m}/S_{60\mu m}\ge 0.18$.
This fraction (20/161=12\%) can be interpreted as an upper limit to
the fraction of FIR-luminous galaxies that may be powered primarily by
an AGN. Genzel et al. (1998) found that an energetically dominant 
active nucleus is present in 20 to 30\% of all FIR-ultraluminous
galaxies based on the analysis of their mid-IR diagnostic diagram.
This broad agreement is not surprising since both methods 
utilize the enhanced mid-IR emission as an indicator.  
The actual fraction of FIR-luminous galaxies powered by 
AGNs may be smaller still as discussed below.  The total
fraction of radio-excess objects among the mid-IR excess
objects is only 10\% (2 out of 20).  Perhaps fortuitously, 
this fraction is similar to the frequency of radio-loud 
objects among optically identified QSOs \citep{Kellermann89,Hooper95}.  

The proposed separation of AGNs from star forming galaxies
using the $S_{25\mu m}/S_{60\mu m}$ flux ratio is not entirely
supported by Figure~\ref{fig:color2}, however.  Both the 
field galaxies and FIR-luminous galaxies show a broad
general trend towards a higher $S_{25\mu m}/S_{60\mu m}$ flux ratio
with increasing $S_{60\mu m}/S_{100\mu m}$ flux ratio.
This trend is seen more clearly when each class of objects are 
considered separately, and this follows the expected behavior
for an aging and evolving starburst.  To quantify this trend,
the color evolution track for a starburst galaxy computed by  
Efstathiou et al. (2000) is shown as a thick curve in Figure~\ref{fig:color2}.  
In this model, a starburst at its onset (top right end of the track) 
displays significant mid-IR emission contributed by the hot/warm
dust directly exposed to the intense radiation field of the young
massive stars.  The predicted $S_{25\mu m}/S_{60\mu m}$
ratio is comparable to the largest values observed for the FIR-luminous 
galaxies.  As the starburst ages and the hot/warm dust cools, the
starburst travels down along the track as shown in Fig.~\ref{fig:color2}.
It is particularly interesting that this color evolution
track closely hugs the proposed AGN separation line at
$S_{25\mu m}/S_{60\mu m}= 0.18$ and outlines the right side
boundary for the FIR-luminous objects in this plot.  The treatment
of the supernova ejecta in this model is such that the evolution
track shown is essentially the right boundary for a family of possible
evolution tracks, in agreement with the data shown here 
-- see Efstathiou et al. (2000) for a detailed description of the model.  
The youngest starbursts ($\le20$ Myr) are predicted to appear well 
to the right of the AGN separation
line, and this prediction is further supported by the fact that
young starburst galaxies hosting 
massive Wolf-Rayet clusters such as NGC~1614, NGC~1741, 
and NGC~5253 show Seyfert-like mid-IR excess with 
$S_{25\mu m}/S_{60\mu m}$ flux ratios of 0.22, 0.15, and 0.39,
respectively, while they show no evidence for hosting an AGN.
Therefore some caution should be taken for interpreting  
mid-IR excess as an AGN indicator, especially 
for galaxies with warm far-IR color.

The fact that an overwhelming majority of high-luminosity objects in
our IRAS selected sample {\it do not} show mid-IR excess and have the
same infrared color characteristics as the less luminous normal star
forming galaxies would argue that most of them are powered by an
intense starburst instead.  The derived radio LF for the
high-luminosity objects in the IRAS 2 Jy sample being not consistent
with that of the radio AGNs or ``monsters" in the UGC sample of Condon
(1989) also indicate that the majority of them may not be powered by
an AGN.  Further, about 30\% (7/23) of the radio-excess objects appear
to the left of the mid-IR excess line in
Figure~\ref{fig:color2}, indicating that some of the radio AGNs do not
contribute significantly to their host's infrared luminosity.

\subsection{Nature of FIR-Excess Galaxies}

There are nine objects in the IRAS 2 Jy sample with
unusually large FIR/radio flux ratios with $q\ge3$.
Three objects have $S_{60\mu m}/S_{100 \mu m}>1$ 
($T_d=60\sim80$ K), and the dust 
temperatures in these galaxies are among the highest in the 
sample.  The strong temperature dependence in dust emissivity
implies that a dramatic increase in FIR luminosity can result 
from only a slight increase in dust temperature.    
Similarly large $S_{60\mu m}/S_{100\mu m}$ ratios are also found 
in infrared quasars FSC~09105+4108 \citep{Hutch88}, 
FSC~13349+2438 \citep{Bei86}, FSC~15307+3252 
(Cutri et al. 1994) and other luminous 
AGN host galaxies (see Yun \& Scoville 1998).  Ironically,
broad emission lines indicative of a luminous AGN have not
been detected by optical or infrared spectroscopy among 
the FIR-excess galaxies observed to date.  Evidence for
large visual extinction and/or large gas and dust contents
is present in the four best studied galaxies (NGC~4418, 
IRAS~08572+3915, IRAS~20087$-$0308, IRAS~22491$-$1808).

Using the high resolution imaging of 8.4 GHz radio emission
in the 40 ultraluminous galaxies in the IRAS Bright Galaxy Sample,
Condon et al. (1991b) have shown that starburst regions in 
some of these galaxies are compact and dense enough to be
opaque even at mid-IR and long radio wavelengths.  Such compact
and intense starbursts can indeed account for the large
$S_{60\mu m}/S_{100 \mu m}$ ratios and high extinction at the same 
time.  These infrared excess objects may actually represent a
special phase in the starburst evolution.

The existing data, however, cannot uniquely distinguish whether these
FIR-excess objects are powered by a dust enshrouded 
AGN or by a compact starburst.  If a dense environment
and associated large opacity is responsible for the observed
infrared excess, then one cannot possibly probe the
energy source directly since the central active region must be 
opaque -- even for hard X-ray photons if the
line of sight column density exceeds $10^{24}$ cm$^{-2}$.
If free-free absorption is responsible for the low 1.4 GHz flux, then
dense ionized gas must fill most of the 1.4 GHz source volume 
($>50-100$ pc in size).  Since the UV opacity is so high, an
ionizing photon cannot travel very far before
being absorbed by dust.  Thus an extended, distributed source 
(a starburst) is favored over a single point source 
(an AGN) to account for the observed ionization.  

What is more certain is that such a compact starburst phase or
a dust-enshrouded AGN phase must be 
brief, perhaps only a few percent of the IR bright phase,
because only nine out of 1809 sources show such infrared 
excess.\footnote{Although this cutoff is somewhat arbitrary, 
they represent the most extreme cases among 
the galaxies responsible for the symmetric dispersion in 
Figure~\ref{fig:qplot}, and such infrared-excess objects
are extremely rare.}  Converting significant fraction 
of the AGN luminosity into thermal dust emission requires a 
geometry for thoroughly enshrouding the AGN with dust.
The paucity of FIR-excess objects in the IRAS 2 Jy sample
thus argues strongly that FIR-luminous galaxies {\it powered} by 
luminous AGNs account for a very small fraction.  The
AGN-powered FIR-luminous galaxies may be in the radio-excess phase
instead, but we have already established in \S~4.2 that the
radio-excess fraction is less than a few per cent, independent of 
luminosity.   

\bigskip

\subsection{Evidence for Non-linearity in the Radio-FIR Correlation}

As a further test of the radio-FIR correlation, the IRAS 60 $\mu$m LF 
is plotted on top of the 1.4 GHz radio LF in Figure~\ref{fig:compareLF} 
after shifting the 60 $\mu$m LF along the x-axis using the linear 
radio-FIR relation in Eq.~4.  
The agreement is extremely good for the two LFs over a wide range 
of luminosity.  A total of 1467 galaxies (81\%) belong to the luminosity 
range between $10^{21}$ and $10^{23}$ W Hz$^{-1}$ where the two LFs are
essentially identical.  
The luminosity range over which the two LFs agree within a factor
of two includes over 98\% of the total sample.  Thus, the similarity 
between the two LFs stems from the fact that
the overwhelming majority of galaxies in our IRAS selected sample
follow the same linear radio-FIR correlation.

Evidence for diminished radio emission among low-luminosity 
galaxies was discussed earlier (see \S3.1 and in Fig.~\ref{fig:qplot}), 
but its collective impact on the radio and FIR LF appears to be minimal.  
This non-linearity among the low-luminosity galaxies should 
result in flattening of the radio LF with respect to the 
60 $\mu$m LF.  Indeed a hint of such an offset is seen 
in Figure~\ref{fig:compareLF}, but the difference is marginal.   
A more substantial difference is seen at the high 
luminosity end, where a dramatic increase in the radio LF over the 
60 $\mu$m LF is seen near $L_{1.4GHz}\sim10^{23.5}$ W Hz$^{-1}$
($L_{60\mu m}\sim 10^{11.5} L_\odot$).  While this may be interpreted 
as a sign of increased AGN activity associated with the high 
luminosity population, the falloff in the LF by an order of magnitude 
near $L_{1.4GHz}\sim10^{24}$ W Hz$^{-1}$ is not consistent with 
such a supposition (see \S5.2).  
The frequency of radio-excess objects was found to be independent 
of luminosity in \S4.2, and the constant AGN fraction 
should appear as a small and constant offset between the two LFs.  
A plausible explanation for the large scatter may be 
the small number statistics in the total numbers of objects 
and the AGN frequency in these bins.  

\medskip

\subsection{FIR Luminosity Density and Local Star Formation Rate}

The derivation of the integrated IRAS 60 $\mu$m luminosity
density for the local volume is shown in Figure~\ref{fig:sumLF60}.  
The derived 60 $\mu$m luminosity density integrated between 
$L_{60\mu m}=10^8 L_\odot$ and $10^{12} L_\odot$ 
for the IRAS 2 Jy sample is $2.6\pm0.2 \times 10^7 L_\odot$ Mpc$^{-3}$.
This estimate lies midway between the values derived from the
60 $\mu$m LFs of Soifer et al. (1987) and ``S17" of
Saunders et al. (1990).  These differences arise mainly from the
different behaviors of these LFs at low luminosity end 
(see Fig.~\ref{fig:60LF}b and \S5.3), and the observed scatter
can be interpreted as a measure of uncertainty resulting from
the sample selection and clustering of galaxies.
In all three cases, the integrated luminosity density reaches
the asymptotic peak value near $L_{60\mu m}=10^{11} L_\odot$, and 
the {\it infrared luminous galaxies 
($L_{FIR}\ge 10^{11} L_\odot$) contribute less than
$10\%$ to the FIR luminosity density in the local volume ($z\le 0.15$)}.

Using Eq.~3 and assuming a median IRAS 60 $\mu$m to 100 $\mu$m flux 
ratio of 0.5 (see Figure~\ref{fig:color2}), we estimate a local 
FIR luminosity density of $(4.8\pm0.5) \times 10^7 L_\odot$ Mpc$^{-3}$.
This is midway between the values derived by Soifer et al. 
($\sim9 \times 10^7 L_\odot$ Mpc$^{-3}$) and
Saunders et al. ($4.2 \times 10^7 L_\odot$ Mpc$^{-3}$).
The derived FIR luminosity density can then be used to derive
the {\it extinction-free} star formation rate (SFR) for the local 
volume assuming the observed FIR luminosity is proportional to the 
level of star forming activity.  

A wide range of conversion relations
between SFR and $L_{FIR}$ are found in the literature, based on either
a starburst model or an observational analysis (e.g., Hunter et al. 
1986, Meurer et al. 1997, Kennicutt 1998ab). 
Using the continuous burst model of 10-100 Myr duration by 
Leitherer \& Heckman (1995) and Salpeter IMF with mass limits of 
0.1 and 100 $M_\odot$, Kennicutt (1998ab) derives a conversion 
relation appropriate for starbursts as
$${SFR}~(M_\odot yr^{-1}) = 1.7 \times 10^{-10} L_{8-1000\mu m}(L_\odot) \eqno(13)$$ 
with about 30\% uncertainty.  The ratio between $L_{8-1000\mu m}$
and $L_{FIR}$ ($\equiv L_{40-120\mu m}$) for the 60 IRAS BGS 
galaxies studied by Sanders et al. (1991) ranges between 1.15-2.0
with a median value of about 1.30.  Based on 
actual measurements at both long and short wavelengths, 
Meurer et al. (1999) find the ratio is larger than 1.4 while
Calzetti et al. (2000) find an average value of 1.75 for the 
five galaxies they examined in detail.  Adopting a conservative ratio 
of 1.5 and using Eq.~13, we
derive the star formation rate of $0.012\pm0.004$ $M_\odot$ yr$^{-1}$ 
Mpc$^{-3}$ for the local volume ($z\simless 0.15$).
Since more than 90\% of the
local FIR luminosity density is contributed by normal, late-type field
galaxies rather than by the starburst population, a conversion
relation derived for late-type field galaxies may be more 
appropriate for the IRAS 2 Jy sample.  The local star
formation density derived using the conversion relation 
derived for the field galaxies by Buat \& Xu (1996)
is slightly larger, $0.015\pm0.005$ $M_\odot$ yr$^{-1}$
Mpc$^{-3}$. Contribution from cold cirrus and smaller optical depth in
late type galaxies add some uncertainty to the inference of 
star-formation rate from FIR luminosity (see Kennicutt 1998a 
for a detailed discussion).    

The two different methods produce a consistent estimate for the
local star formation density.  In comparison, the the local
SFR derived from the H$\alpha$ luminosity
density measured by Gallego et al. (1995) using the conversion
relation given by Kennicutt (1998a) is $0.014\pm0.007~M_\odot$
yr$^{-1}$ Mpc$^{-3}$.
An important consequence is that the local SFR derived 
from the FIR luminosity density
provides an independent estimate for $z=0$ and lends a further
support for the $z=0$ normalization for the cosmic star-formation 
history models such as those by Blain et al. (1999; dashed line) and by 
Tan et al. (1999; solid line) as well as support for the larger
extinction correction for all UV derived values as proposed by
Steidel et al. (1999) and others (see Figure~\ref{fig:madau}).

The SFR derived from the FIR luminosity density 
may be an overestimate if the observed FIR includes 
a significant contribution from dust heated by an AGN or by 
interstellar radiation field. The frequency of the radio-excess 
objects in our sample is only about 1\%, independent of luminosity.
And the overall frequency of AGNs (including radio-quiet) may  
be as high as 10\% if only one out of ten AGNs produces 
significant radio emission.  
Some AGN contribution to the total FIR luminosity density is undoubtedly 
present, but it is nevertheless not likely to be dominant (see \S6.1 \&
\S6.2).
The contribution from the dust heated by interstellar radiation 
field has been suggested previously to account for up to 50 to 
70\% of the FIR luminosity (see Lonsdale-Persson \& Helou 1987,
Sauvage \& Thuan 1992, Xu \& Helou 1996).  
This claim does not seem entirely
consistent with our finding that an overwhelming
majority of both the field and starburst galaxies obey the same 
linear radio-FIR relation (see \S3 and \S6.2). 
A possible explanation may be that the cirrus is heated
by non-ionizing UV photons from relatively massive stars,
so that even the cirrus tracks the population of stars
producing the radio emission (Xu 1990).  

\medskip

\subsection{1.4 GHz Radio Luminosity and Star Formation Rate}

The excellent linear correlation seen between the FIR and radio 
luminosity among the IRAS 2 Jy sample offers
a possibility of deriving the star formation rate
directly from the measured radio luminosity.
Because of the significant scatter in the radio-FIR relation
and possible non-linear effects discussed in \S3, determining the
star formation rate for any particular galaxy using its
radio luminosity alone may be risky.  The effect of the scatter is 
substantially reduced, however, when the radio LF is integrated
over luminosity in deriving the radio luminosity density.  
Further, the star formation density for a given 
volume can be determined with a greater accuracy if the
integration is limited over the range of radio luminosity 
where the linear radio-FIR relation is secure.

Deriving the star-formation density from the radio continuum
luminosity density for the general galaxy population also has to 
account for the possible contribution from radio AGNs.  
As shown in Figure~\ref{fig:sumLF14}, the 1.4 GHz radio LF is dominated 
by the normal late type galaxies for $L_{1.4GHz}\le 10^{23}$
W Hz$^{-1}$ while the contribution by the ``monsters" becomes
significant only at $L_{1.4GHz}\ge 10^{24}$ W Hz$^{-1}$. This 
bimodality is also clearly seen in the histogram of 1.4 GHz radio luminosity 
distribution for Condon's UGC sample shown in Figure~\ref{fig:L14histo}.
Since the net AGN contribution is small and quantitatively well
understood for $L_{1.4GHz}\le 10^{24}$ W Hz$^{-1}$, the star formation 
density for the local volume can be derived with some confidence from the
radio luminosity density integrated up to this luminosity cutoff.  

The integrated 1.4 GHz luminosity density derived from the IRAS
2 Jy sample for $L_{1.4GHz}\le 10^{24}$ W Hz$^{-1}$ is
$2.6 \times 10^{19}$ W Hz$^{-1}$ Mpc$^{-1}$ (see Figure~\ref{fig:sumLF14}).  
The empirical conversion relation that links this 1.4 GHz luminosity 
density to the local star formation density of 
$0.015\pm0.005~M_\odot$ yr$^{-1}$ is then
$${SFR}~(M_\odot yr^{-1}) = 5.9\pm1.8 \times 10^{-22} L_{1.4GHz}(W~Hz^{-1}). \eqno(14)$$

As discussed in \S6.4, the main uncertainty in this relation 
comes from the estimate of the local star formation density.
If the radio LF at a given epoch
can be derived with sufficient accuracy, then the 
contribution by star forming galaxies and radio AGNs can be 
distinguished and the star formation density for the epoch 
can be derived using Eq.~14.  Whether the dividing line between
the starburst and AGN population near $L_{1.4GHz}\sim 10^{23.5}$ 
W Hz$^{-1}$ is constant or evolved over time is unknown at
the moment, and it needs to be investigated by future studies.
Ultimately, the cosmic star formation 
history free of any extinction effects may be derived using 
the radio emission from star forming galaxies and by tracking 
the evolution in their radio luminosity function.

\section{Conclusions}

By cross-correlating the IRAS Redshift Survey catalog and the NRAO VLA
Sky Survey database, we have assembled a flux-limited complete sample
of 1809 IRAS detected galaxies.  This IRAS 2 Jy sample is 6 times
larger in size and 5 times deeper in redshift coverage ($z\le 0.15$)
than previous studies of radio and infrared properties of galaxies.
In addition to obtaining the accurate positions of these IRAS sources,
we have examined the well known radio-FIR correlation in detail.  The
principal conclusions from the analysis of the data are the following:

1. The radio-FIR correlation for the FIR-selected
galaxy sample is well described by a linear
relation, and over 98\% of
the sample galaxies follow this linear radio-FIR correlation.
The scatter in the linear relation is about 0.26 in dex, dominated
by small systematic deviations involving a small number of galaxies 
at high and low luminosity ends.

2. A total of 23 ``radio-excess" objects (radio AGNs) are found in
our sample including three ``radio-loud" objects PKS~1345+12, 3C~273,
and NGC 1275.  The frequency of radio-excess objects in our IRAS 2 Jy
sample is 23/1809 = 1.3\%, and radio-excess objects are rare among the
FIR-selected sample of normal and starburst galaxies.  There are only
3 radio-excess objects among the 161 FIR-luminous ($L_{60\mu
m}\ge10^{11.3} L_\odot$) galaxies in the sample (3/161 = 1.9\%).
Therefore the frequency of radio AGNs appears to be independent of FIR
luminosity.

3. The derived 1.4 GHz radio luminosity function and 60 $\mu$m 
luminosity function for the IRAS 2 Jy sample are both reasonably 
modeled as linear sums of two Schechter functions, one representing 
late type field galaxies and the second representing a starburst 
population.  The presence of the second, high luminosity component 
in these LFs is a predictable outcome of the elevated M/L associated
with the starburst population, and the Press-Schechter formulation 
of the structure formation can now be naturally extended to the 
interpretation of the radio and infrared LFs as well.

4. The radio-excess objects in our sample show evidence for
the same highly enhanced radiation field as the infrared luminous
galaxies in the infrared color-color diagrams.  A systematic 
segregation of field galaxies, starburst galaxies, and radio 
AGNs is found in the mid-IR versus far-IR color-color diagram 
(Fig.~\ref{fig:color2}), where the radio AGNs show the largest 
observed $S_{25\mu m}/S_{60\mu m}$ ratios.  
The total fraction of radio-excess objects among the galaxies 
with mid-IR enhancement is only about 10\% (2/20) -- similar to 
the frequency of radio-loud objects among the QSOs.  The absence 
of mid-IR enhancement and radio-excess among the large majority 
of the IR luminous galaxies 
suggests that star formation is the dominant energy source in most cases.

5. A small number of infrared-excess objects are also identified by
their departure from the radio-FIR relation (9 objects with
$q\ge3$). Both dust-enshrouded AGNs and compact starbursts can explain
their observed properties, and distinguishing the two may be
difficult.  The observed spatial extents of 50-100 pc favors the
starburst explanation.  In either case, this phase must be brief, 
perhaps lasting only a few percent of the IR luminous phase since 
these infrared-excess objects are extremely rare.

6. The integrated 60 $\mu$m luminosity density between 
$L_{60\mu m}=10^8 L_\odot$ and $10^{12} L_\odot$ 
is $2.6\pm0.2 \times 10^7 L_\odot$ Mpc$^{-3}$.
The FIR-luminous galaxies 
($L_{FIR}\ge 10^{11} L_\odot$) contribute less than 10\% of
the total.  The derived local FIR luminosity density is  
$(4.8\pm0.5) \times 10^7 L_\odot$ Mpc$^{-3}$.
The inferred {\it extinction-free} star-formation density for the 
local volume is $0.015\pm0.005~M_\odot$ yr$^{-1}$ Mpc$^{-3}$.  

7. The observed linear correlation between the FIR and radio 
luminosities in the IRAS 2 Jy sample offers
a potentially useful method of deriving the star-formation rate
using the measured radio luminosity density.  
A good separation found between star-forming galaxies
and AGNs in the derived radio LF offers a promising possibility
of tracing the evolution of the two populations separately
by deriving their radio LFs at different epochs.

By examining the radio properties of a large sample of 
infrared selected galaxies, we have obtained a much better
understanding of the extent of the radio-FIR correlation.
We have also obtained a good quantitative understanding
of the frequency and energetic importance of active galactic 
nuclei among the infrared selected galaxies and outlined the
practical limits on using the radio continuum as a direct
probe of star forming activity.  This study can now serve
as the basis for conducting deep radio, IR/submm, and optical surveys 
aimed at mapping the extinction-free cosmic-star formation history
and the evolution of radio AGNs.

\acknowledgements

The authors are grateful to W. Cotton, F. Owen, D. Sanders, 
P. Schechter, N. Z. Scoville, C. Xu,
and many others for insightful discussions.   
M. Yun was supported by the NRAO Jansky Fellowship during the
initial phase of this project. N. Reddy was
supported by the Summer Student Research Assistantship 
provided by the National Science Foundation and the National
Radio Astronomy Observatory.  
The National Radio Astronomy Observatory is a facility of the 
National Science Foundation
operated under cooperative agreement by Associated Universities, Inc.
Some of the data presented here are obtained from the
NASA/IPAC Extragalactic Database (NED), which is operated by the Jet
Propulsion Laboratory, California Institute of Technology, under
contract with the National Aeronautics and Space Administration.

\clearpage

\begin{figure}
\plotone{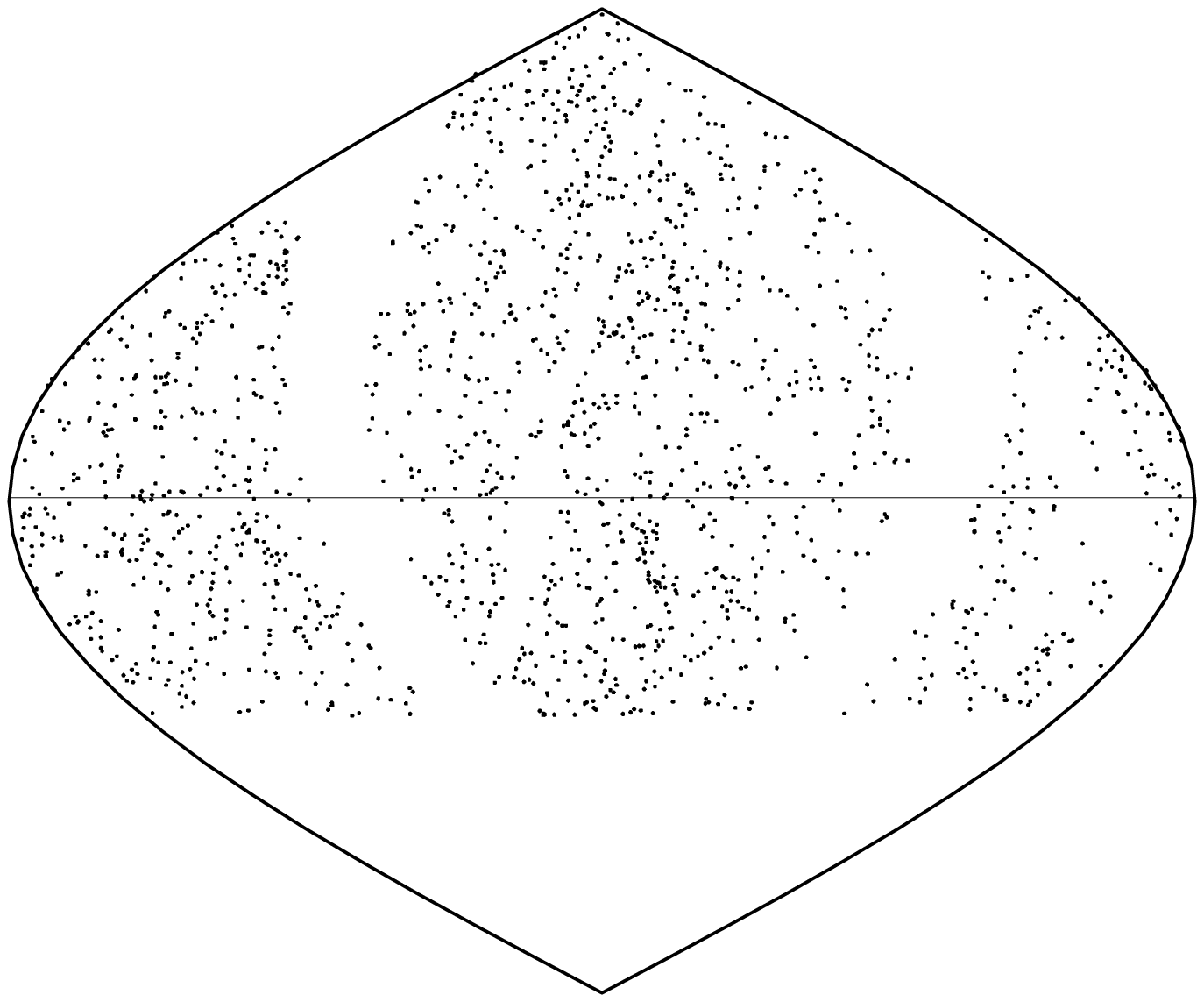}
\caption{
Equal area sky projection of the IRAS 2 Jy 
Sample.  Large gap to the right and left of the center
trace the Galactic Plane, and a narrow empty strip near the center
is the well known gap in the IRAS survey.  Otherwise, the sky coverage
is fairly uniform.  \label{fig:skyplot}}
\end{figure}

\begin{figure}
\plotone{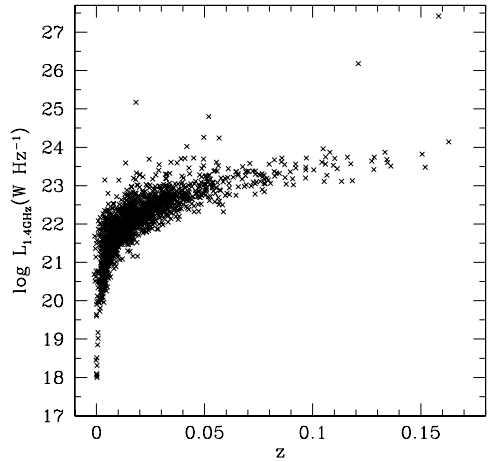}
\caption{
Plot of 1.4 GHz radio luminosity versus redshift 
for the IRAS 2 Jy sample.  The lower bound is defined by the 2 Jy 
flux limit of the IRAS 60 $\mu$m band.  The upper bound is dictated 
by the radio luminosity function.  The three galaxies with 
$L_{1.4GHz}\ge10^{25}$ W Hz$^{-1}$ are identified as ``radio-loud" 
objects.  \label{fig:lzplot}}
\end{figure}

\begin{figure}
\psfig{file=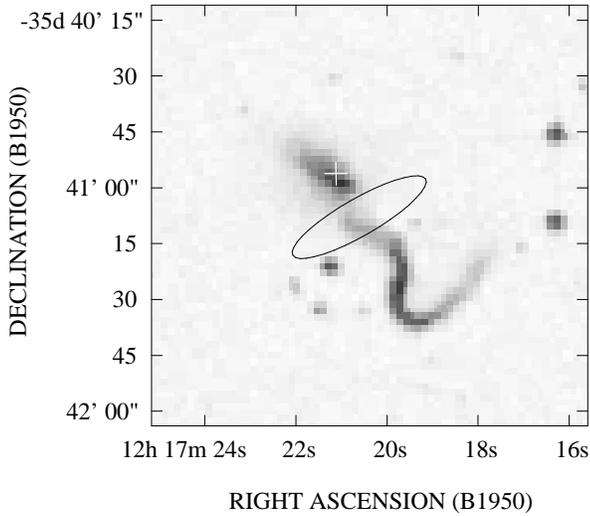,width=3.0in,angle=270}
\caption{
An example of confusing source identification
using the NVSS catalog.  The IRAS PSC position for IRAS~12173$-$3541
and its error ellipse are centered between the two interacting galaxies 
while the dominant radio source is identified with the northern 
galaxy by the NVSS (marked with a white cross).  \label{fig:i12173}}
\end{figure}

\begin{figure}
\plotone{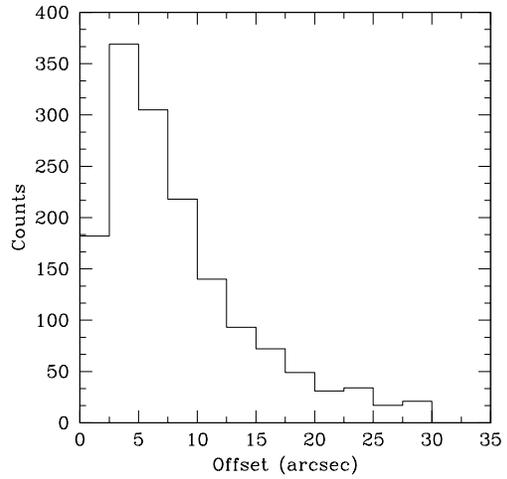}
\caption{
A histogram of the offsets in arcsec between the IRAS PSC and 
the NVSS catalog positions. 
The distribution is not Gaussian because the IRAS position error
ellipse is typically highly elongated. 
The position offset is less than 10\arcsec\ in over 2/3 of the 
cases, which is consistent with the expectation from the 
IRAS position uncertainty.  \label{fig:offsets}}
\end{figure}

\begin{figure}
\psfig{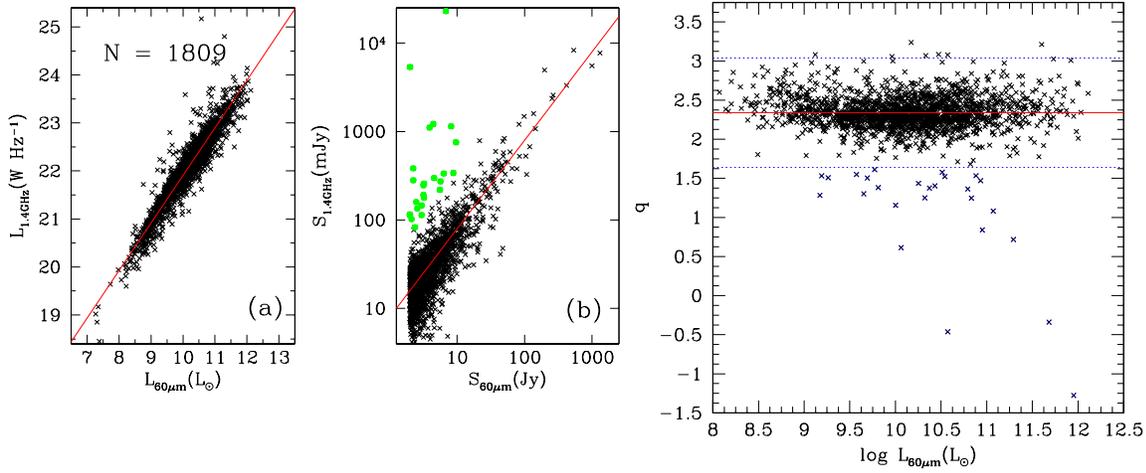}
\caption{
(a) Plot of 1.4 GHz radio luminosity versus IRAS 60 $\mu$m luminosity. 
The solid line corresponds to a linear relation with a constant offset.
(b) Plot of 1.4 GHz and IRAS 60 $\mu$m flux density for the IRAS
2 Jy sample.  The solid line corresponds to the same linear relation 
shown in (a).  Gray filled circles identify the radio-excess 
objects (see \S4.2). The remaining 
$\sim$1750 objects (out of 1809) lie very close to the linear relation, 
and the rms scatter of the data is less than 0.26 in dex.
\label{fig:lums}}
\end{figure}

\begin{figure}
\plotone{f6.cps}
\caption{
Distribution of $q$ values plotted as a function of IRAS 60 $\mu$m 
luminosity.  The solid line
marks the average value of $q=2.34$ while the dotted lines delineate the
``radio-excess" (below) and ``infrared-excess" (above) objects,
delineated for having 5 times larger radio and infrared flux density than
the expected values from the linear radio-FIR relation, respectively.
\label{fig:qplot}}
\end{figure}
\newpage

\begin{figure}
\vspace{-35mm}
\psfig{file=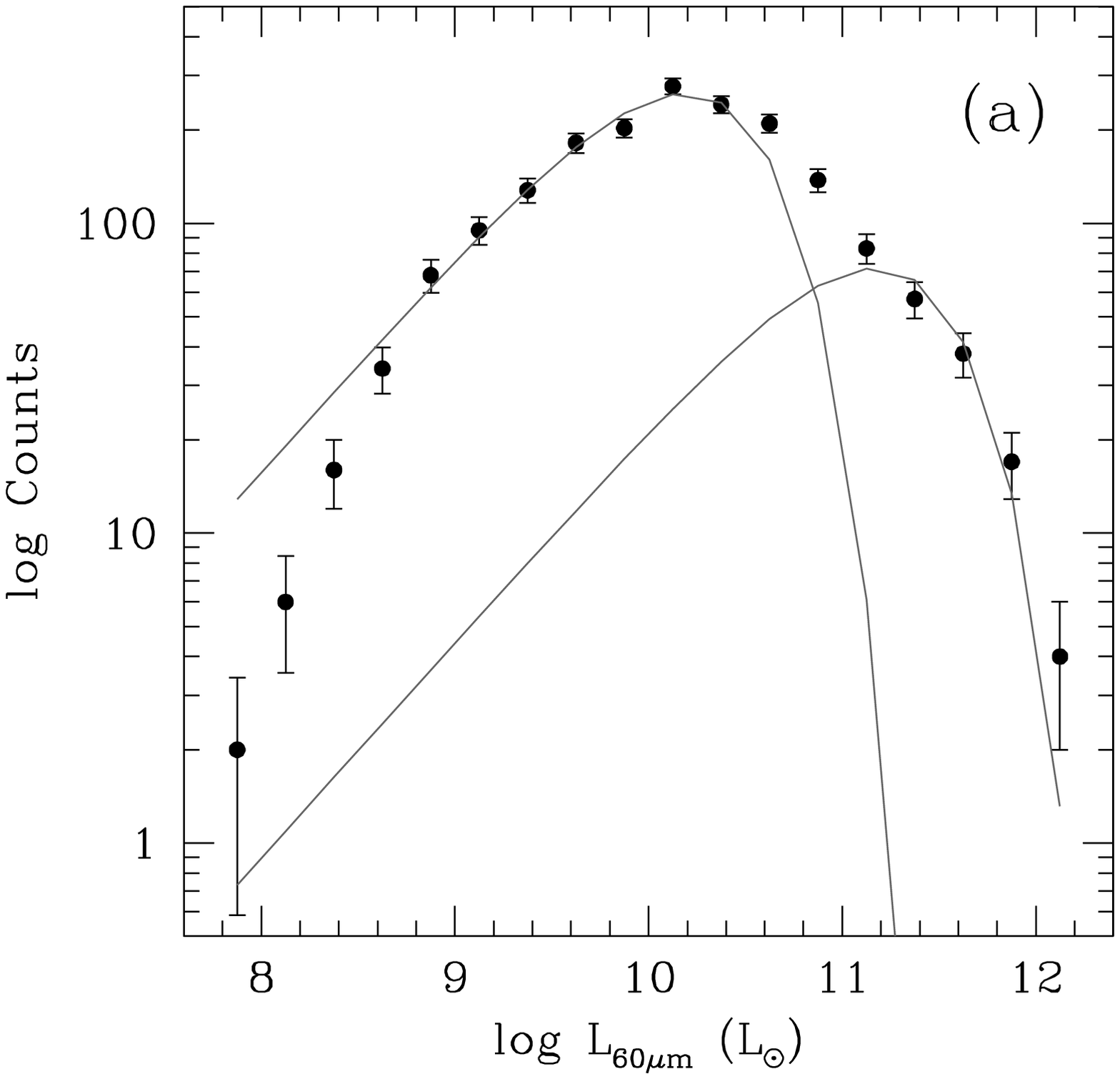,width=2.5in,angle=0}
\vspace{-12mm}
\psfig{file=f7b.cps,width=2.5in,angle=0}
\vspace{-12mm}
\psfig{file=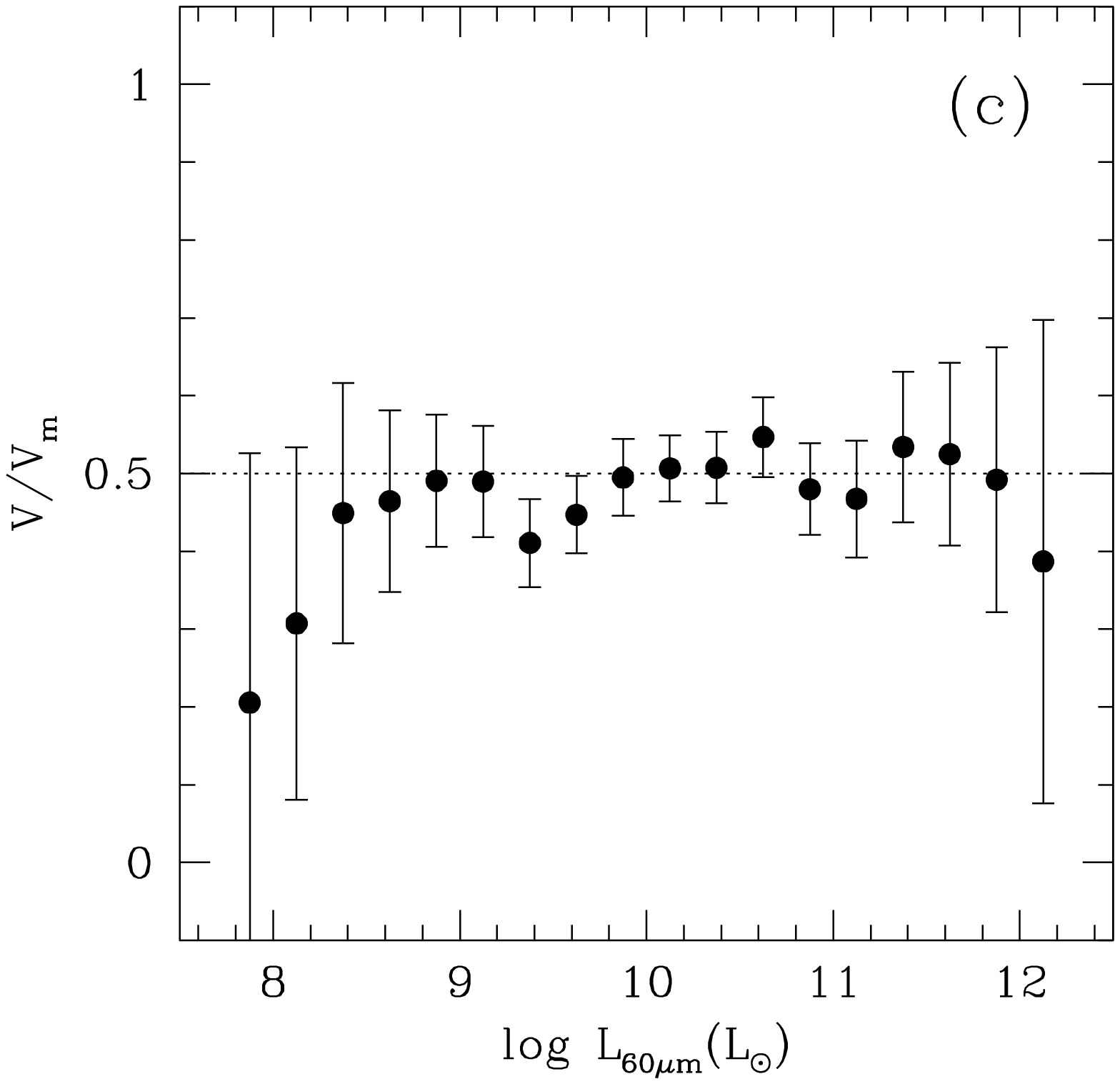,width=2.5in,angle=0}
\vspace{-10mm}
\caption{
(a) 60 $\mu$m luminosity distribution for the IRAS 2 Jy sample.  
The data points are number of counts in each
luminosity bin, and the error bars reflect the Poisson statistics, 
$N^{1/2}$ (see Table~4).  The two solid lines indicate the 
best-fit curves of the Schechter form listed in Table~5. 
(b) The $60\mu$m infrared luminosity function.  Filled circles
represent the LF for the IRAS 2 Jy sample.  The open squares
and open circles represent the 60 $\mu$m LFs derived from the
IRAS BGS sample by Soifer et al. (1987) and an ensemble of
data (``S17") analyzed by Saunders et al. (1990), respectively.
The long dashed lines correspond to the two Schechter fits
for the high and low luminosity objects listed in Table~5.
The solid line is the resulting linear sum of the two 
Schechter functions.  The dotted curve represents the
double power-law LF derived from an earlier version of the
IRAS redshift survey sample by Yahil et al. (1991) using a
density-independent determination method.
(c) The $V/V_m$ plot for the IRAS $60\mu$m
luminosity distribution.  The $V/V_m$ analysis suggests that the highest
and the lowest luminosity bins are not well sampled and may
suffer a completeness problem.
\label{fig:60LF}}
\end{figure}
\newpage

\begin{figure}
\vspace{-35mm}
\psfig{file=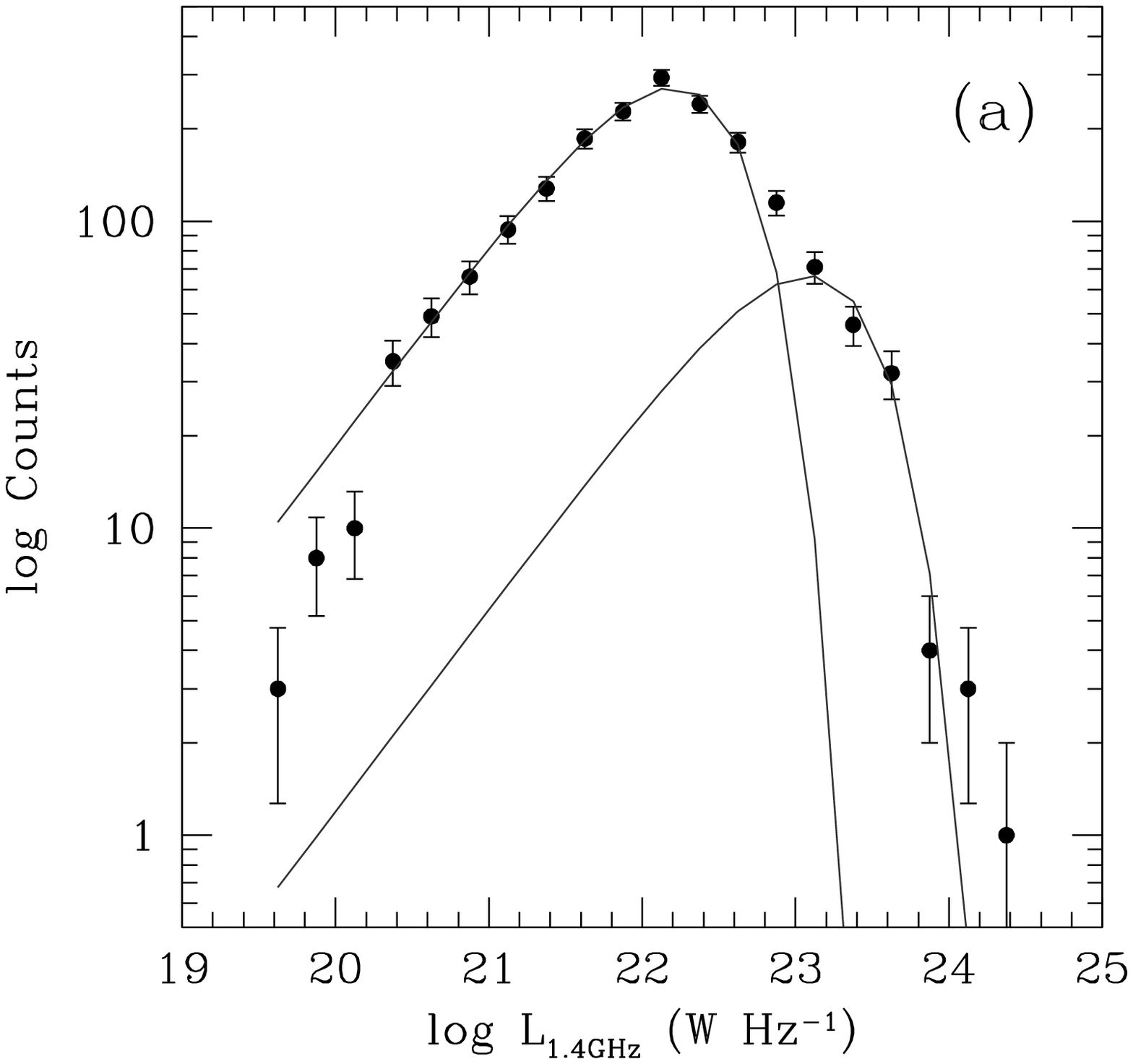,width=2.5in,angle=0}
\vspace{-12mm}
\psfig{file=f8b.cps,width=2.5in,angle=0}
\vspace{-12mm}
\psfig{file=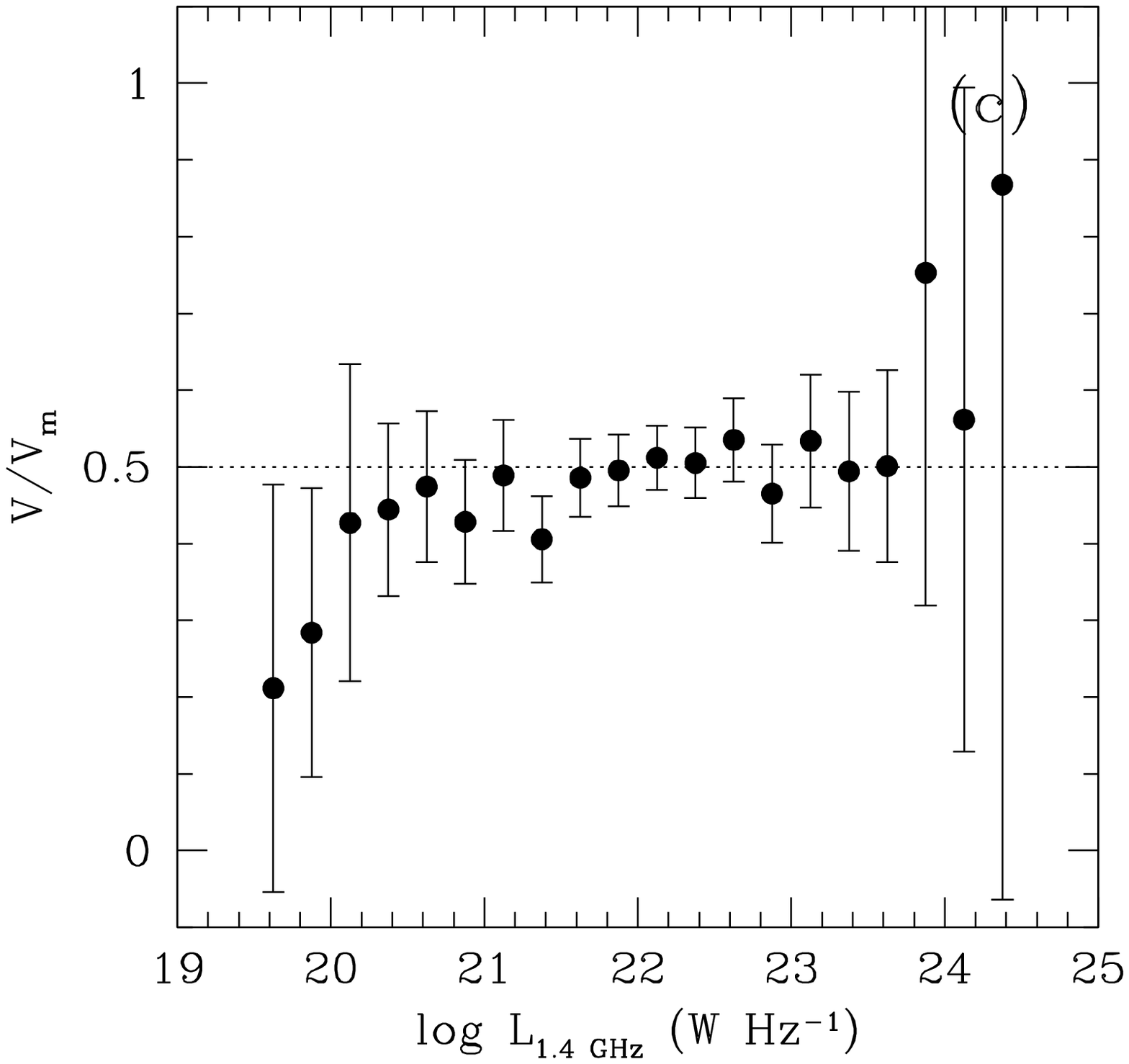,width=2.5in,angle=0}
\vspace{-5mm}
\caption{
(a) 1.4 GHz radio luminosity distribution for the
IRAS 2 Jy sample.  The data points are number counts in each
luminosity bin, and the error bars reflect the Poisson statistics, 
$N^{1/2}$ (see Table~6).  The two solid lines indicate the 
best-fit curves of the Schechter form listed in Table~7. 
(b) The 1.4 GHz radio luminosity function.  Filled circles
represent the IRAS 2 Jy sample while the open circles and
squares represent the 1.4 GHz LFs for the ``starburst" and 
``monsters" from Condon (1989), respectively.  The solid line is the best
fit Schechter function for the low luminosity objects only
(see Table~7).
(c) The $V/V_m$ plot for the 1.4 GHz radio
luminosity distribution.  The error bars are statistical 
and may not be accurate for bins with small number of counts.
The $V/V_m$ analysis suggests that the highest
and the lowest luminosity bins are not well sampled and may
suffer a completeness problem.
\label{fig:1.4LF}}
\end{figure}

\begin{figure}
\plotone{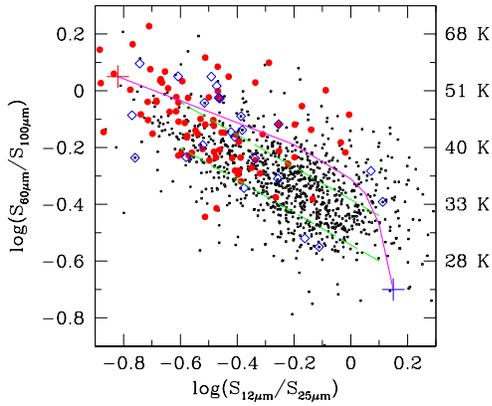}
\caption{
IRAS mid-IR to far-IR color-color diagram for our IRAS 2 Jy
Sample.  Large filled circles correspond to infrared luminous
galaxies with $L_{60\mu}\ge 10^{11.3} L_\odot$ (see Table~1). 
The low luminosity, late type galaxies in the field
are plotted in small dots.  The large diamonds identify
the ``radio-excess" objects (radio AGNs; Table~2).  The twin broken
parallel lines trace the area occupied by ``normal galaxies" as
identified by Helou (1986), and the curved solid line is the 
theoretical trajectory of mixed composition dust under increasing
mean radiation field calculated by D\'{e}sert (1986), from solar
neighborhood intensity (``cirrus", lower right) to several
hundred times larger radiation field (upper left).
\label{fig:color}}
\end{figure}

\begin{figure}
\plotone{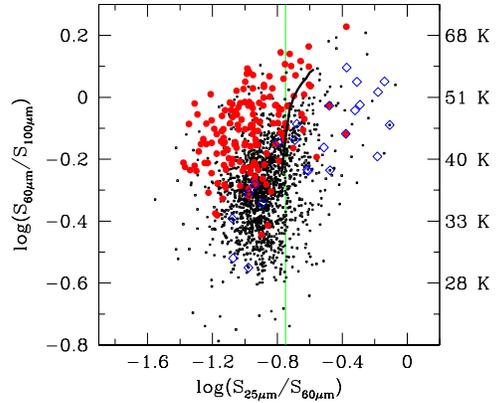}
\caption{
IRAS far-IR versus mid-IR color-color plot for the IRAS 2 Jy 
sample.  The $S_{60\mu m}/S_{100\mu m}$ ratio is translated
into dust temperature assuming emissivity index 
$\beta=1$ (see Helou et al. 1988).  
All the symbols are the same as in Figure~\ref{fig:color}.  
The thin vertical line marks the proposed division
for the infrared Seyferts ($S_{25\mu m}/S_{60\mu m} \ge 0.18$)
by de Grijp et al. (1985).  The thick solid line represents the
color evolution track for a model starburst galaxy from
Efstathiou et al. (2000), starting at zero age on the top right corner
to 72 Myr old starburst at the end (see their Fig.~2 for more detail).
\label{fig:color2}}
\end{figure}

\begin{figure}
\plotone{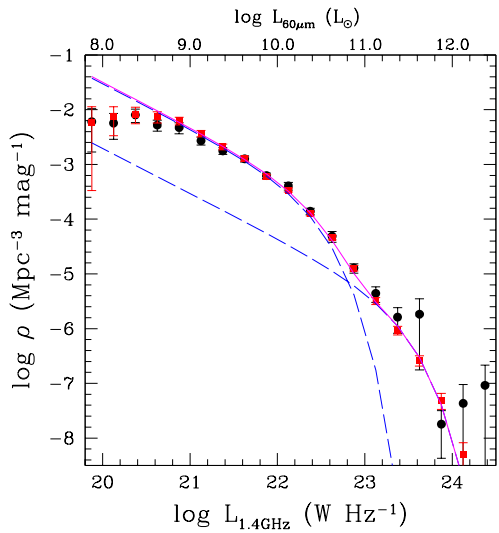}
\caption{
The 60 $\mu$m FIR LF (in filled squares) is translated using the 
linear radio-FIR relation in Eq.~4 and compared directly with the 
1.4 GHz radio LF (in filled circles).  The two Schechter functions
for the high and low luminosity 60 $\mu$m FIR LFs are shown in
long dashed lines and the sum of the two are shown in thick
solid line as in Figure~\ref{fig:60LF}b.  
\label{fig:compareLF}}
\end{figure}

\begin{figure}
\plotone{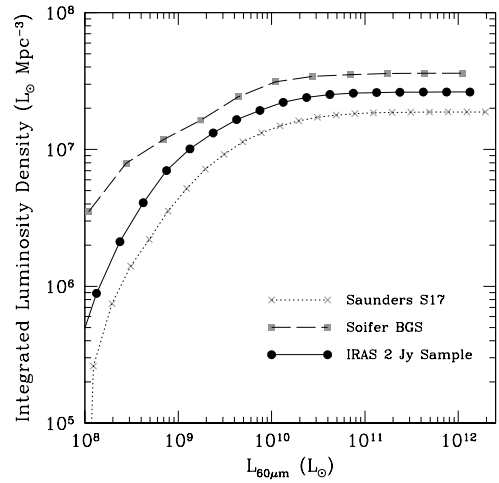}
\caption{
A plot of the cumulative IRAS
60 $\mu$m luminosity density.  The filled circles mark the 
IRAS 2 Jy sample while the filled squares and crosses represent
the IRAS BGS sample (Soifer et al. 1987) and the sample ``S17"
of Saunders et al. (1990).  
\label{fig:sumLF60}}
\end{figure}

\begin{figure}
\psfig{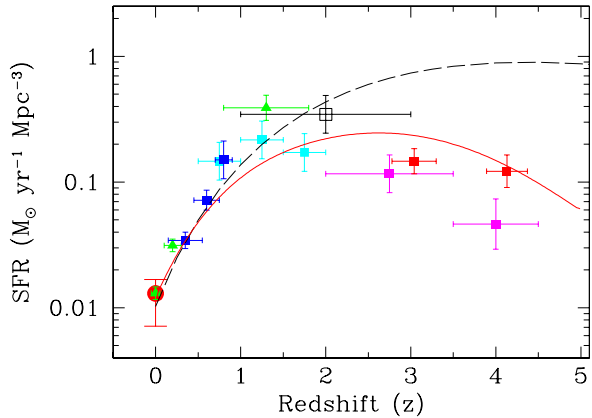}
\caption{
Plot of the cosmic star formation history.
The filled triangles and squares represent the SFRs derived 
from the H$\alpha$ (Gallego et al. 1995, Tresse \& Maddox 1998, 
Yan et al. 1999) and UV continuum (Lilly et al. 1996, Madau et al. 1996, 
Connolly et al. 1997, Steidel et al. 1999) luminosity density, respectively.
All UV density derived values are scaled up by a factor of three to
correct for extinction (see Steidel et al. 1999).  The large empty
square represents the SFR estimated for the faint submm population
at $z=1\sim3$ (Barger et al. 1999).   The large filled
circle represents the local star formation density 
derived from the integrated FIR luminosity density, which agrees 
very well with the SFR derived from H$\alpha$ measurement by 
Gallego et al. (1995).  Two representative models of the
star formation evolution by Blain et al. (1999; dashed line) and 
by Tan et al. (1999; solid line) are also shown after a 
re-normalization to the new estimate of the current star formation rate.
\label{fig:madau}}
\end{figure}

\begin{figure}
\plotone{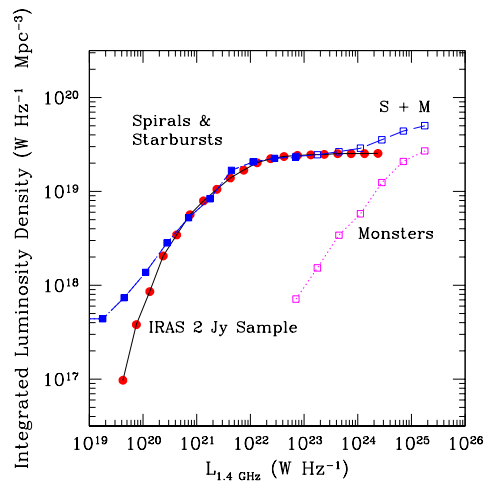}
\caption{
A comparison of the cumulative luminosity
density for the 1.4 GHz radio luminosity functions for the
IRAS 2 Jy sample with those of the ``starbursts \& spirals"
and ``monsters" in the UGC sample studied by Condon (1989).
While radio AGNs with lower luminosity are present, they do
not contribute significantly to the integrated radio luminosity
density below $L_{1.4GHz}\approx 10^{24}$ W Hz$^{-1}$.
\label{fig:sumLF14}}
\end{figure}

\begin{figure}
\plotone{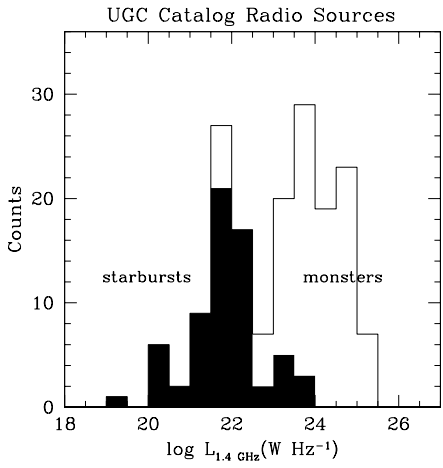}
\caption{
Histogram of 1.4 GHz radio luminosity distribution for
the UGC galaxies studied by Condon (1989).  The 
``starbursts" and ``monsters" show a relatively clean
bimodal separation with a dividing line near
$L_{1.4GHz}\approx 10^{23}$ W Hz$^{-1}$.
\label{fig:L14histo}}
\end{figure}

\clearpage

\begin{deluxetable}{lcccccl}
\tabletypesize{\scriptsize}
\tablecaption{Infrared Luminous Galaxies with log $L_{FIR}\ge 11.3 L_\odot$ \label{tab:catalog}}
\tablehead{
\colhead{Name}              & 
\colhead{$<cz>$}      & 
\colhead{$m$}  &
\colhead{log $L_{1.4GHz} $} &
\colhead{log $L_{60\mu m} $}      &
\colhead{q} & 
\colhead{notes} \\
\colhead{ }   &  
\colhead{km/s}   &  
\colhead{mag}   & 
\colhead{(W~Hz$^{-1}$)}   &
\colhead{$(L_\odot)$} }  
\startdata
IRAS~00057$+$4021 &  13516 & 16.8 & 22.47 & 11.19 & 2.872 & \\
IRAS~00091$-$0738 &  35509 & 18.4 & 23.13 & 11.79 & 2.807 & jet-like extension to south\\
IRAS~00188$-$0856 &  38550 & 18.4 & 23.74 & 11.88 & 2.315 & interacting \\
IRAS~00234$-$0312 &  20219 & 17.0 & 23.17 & 11.40 & 2.409 & VV 837, pair \\
IRAS~00262$+$4251 &  29129 & 16.0 & 23.72 & 11.66 & 2.079 & LINER\\
IRAS~00267$+$3016 &  15119 & 14.8 & 23.24 & 11.28 & 2.230 & MRK 551, blue compact\\
IRAS~00308$-$2238 &  18381 & 17.0 & 23.06 & 11.12 & 2.325 & \\
IRAS~00335$-$2732 &  20771 & 17.5 & 23.02 & 11.54 & 2.633 & starburst\\
IRAS~00456$-$2904 &  33060 & 17.5 & 23.71 & 11.70 & 2.185 & interacting, starburst\\
IRAS~00537$+$1337 &  24673 & 16.5 & 23.37 & 11.35 & 2.206 & \\
IRAS~01003$-$2238 &  35286 & 18.9 & 23.57 & 11.70 & 2.258 & HII/WR, Sy2\\
IRAS~01053$-$1746 &   6016 & 14.3 & 23.29 & 11.19 & 2.084 & IC1623, pair, HII\\
IRAS~01106$+$7322 &  13217 & -- & 22.98 & 11.16 & 2.503 & Spiral \\
IRAS~01160$-$0029 &  14274 & 15.1 & 23.01 & 11.11 & 2.306 & galaxy-pair\\
IRAS~01173$+$1405 &   9362 & 14.9 & 22.97 & 11.25 & 2.416 & HII \\
IRAS~01249$-$0848 &  14592 & 15.5 & 23.23 & 11.12 & 2.142 & MRK 995\\
IRAS~01257$+$5015 &  15522 & -- & 23.09 & 11.02 & 2.220 & Spiral\\
IRAS~01264$+$3302 &  24145 & -- & 23.06 & 11.36 & 2.489 & \\
IRAS~01364$-$1042 &  14250 & -- & 22.84 & 11.22 & 2.703 & LINER \\
IRAS~01418$+$1651 &   8245 & 16.3 & 22.77 & 11.22 & 2.599 & pair, HII+LINER \\
IRAS~01484$+$2220 &   9705 & 13.7 & 23.18 & 11.12 & 2.185 & NGC 695, pec\\
IRAS~01572$+$0009 &  48869 & 15.2 & 24.14 & 12.00 & 2.012 & Sy1\\
IRAS~02153$+$2636 &  15016 & -- & 23.04 & 11.05 & 2.268 & \\
IRAS~02203$+$3158 &  10142 & 14.9 &  23.06 & 11.09 & 2.286 & MRK 1034, pair of Sy1\\
IRAS~02290$+$2533 &  15557 & -- & 23.25 & 11.06 & 2.080 & \\
IRAS~02378$+$3829 &  15052 & 15.7 & 22.94 & 11.11 & 2.369 & pair \\
IRAS~02483$+$4302 &  15571 & -- & 23.02 & 11.29 & 2.502 & QSO-galaxy pair\\
IRAS~02597$+$4455 &  16087 & -- & 22.94 & 11.18 & 2.384 & Spiral \\
IRAS~03119$+$1448 &  23006 & 17.3 & 22.90 & 11.33 & 2.597 &  \\
IRAS~03158$+$4227 &  40288 & 17.7 & 23.68 & 12.11 & 2.587 & \\
IRAS~03503$-$0328 &  21730 & -- & 23.09 & 11.44 & 2.520 & \\
IRAS~03521$+$0028 &  45622 & 18.0 & 23.48 & 11.99 & 2.720 & LINER\\
IRAS~03593$-$3149 &  21462 & -- & 22.99 & 11.29 & 2.497 & \\
IRAS~04154$+$1755 &  16659 & -- & 23.61 & 11.31 & 1.897 & one-sided jet, Sy2\\
IRAS~04232$+$1436 &  23972 & -- & 23.56 & 11.55 & 2.173 & LINER\\
IRAS~05083$+$7936 &  16288 & 15.8 & 23.38 & 11.42 & 2.280 & patchy, compact\\
IRAS~05189$-$2524 &  12760 & 15.5 & 23.01 & 11.59 & 2.718 & Sy2, pec\\
IRAS~05246$+$0103 &  29037 & 17.0 & 23.48 & 11.59 & 2.232 & GPS\\
IRAS~06076$-$2139 &  11226 & -- & 22.77 & 11.16 & 2.583 & \\
IRAS~06102$-$2949 &  18297 & -- & 23.44 & 11.43 & 2.214 & triplet in cluster \\
IRAS~06206$-$3646 &  32390 & -- & 23.87 & 11.62 & 1.919 & \\
IRAS~07045$+$4705 &  19636 & -- & 23.27 & 11.23 & 2.145 & \\
IRAS~08030$+$5243 &  25006 & 17.0 & 23.33 & 11.53 & 2.398 & \\
IRAS~08071$+$0509 &  15650 & -- & 23.28 & 11.32 & 2.244 & radio jet, HII\\
IRAS~08323$+$3003 &  17885 & 18.0 & 22.74 & 11.26 & 2.693 & \\
IRAS~08340$+$1550 &  23414 & -- & 23.05 & 11.31 & 2.459 & \\
IRAS~08344$+$5105 &  29028 & -- & 23.20 & 11.49 & 2.478 & \\
IRAS~08507$+$3520 &  16748 & 15.0 & 23.38 & 11.03 & 1.925 & UGC~4653, triple \\
IRAS~08572$+$3915 &  17480 & -- & 22.50 & 11.61 & 3.218 & LINER, Sy2 \\
IRAS~08573$-$0424 &  18425 & 17.0 & 22.94 & 11.19 & 2.441 & \\
IRAS~08579$+$3447 &  19645 & 16.0 & 23.31 & 11.29 & 2.210 & Spiral \\
IRAS~09014$+$0139 &  16081 & 17.2 & 23.06 & 11.14 & 2.298 & \\
IRAS~09018$+$1447 &  14847 & -- & 23.69 & 11.25 & 1.788 & Mrk~1224, quadruple\\
IRAS~09061$-$1248 &  22073 & 15.8 & 23.30 & 11.51 & 2.417 & \\
IRAS~09111$-$1007 &  16438 & 16.1 & 23.39 & 11.56 & 2.369 & radio double, interacting\\
IRAS~09126$+$4432 &  11773 & 14.9 & 23.04 & 11.17 & 2.364 & UGC 4881, pair, HII\\
IRAS~09192$+$2124 &  23258 & -- & 23.06 & 11.42 & 2.516 & \\
IRAS~09320$+$6134 &  11762 & 15.5 & 23.70 & 11.48 & 2.007 & UGC 5101, merger\\
IRAS~09432$+$1910 &  16048 & -- & 22.96 & 11.24 & 2.452 & \\
IRAS~09433$-$1531 &  15808 & 16.2 & 23.17 & 11.31 & 2.296 & \\
IRAS~09583$+$4714 &  25717 & 16.3 & 23.72 & 11.52 & 1.959 & Sy1+Sy2 \\
IRAS~10035$+$4852 &  19421 & -- & 23.36 & 11.51 & 2.340 & pair \\
IRAS~10039$-$3338 &  10223 & 15.0 & 22.75 & 11.28 & 2.667 & IC~2545, pair \\
IRAS~10173$+$0828 &  14390 & 17.5 & 22.69 & 11.36 & 2.808 & megamaser \\
IRAS~10190$+$1322 &  22987 & 17.5 & 23.31 & 11.50 & 2.413 & \\
IRAS~10311$+$3507 &  21274 & 15.0 & 23.58 & 11.34 & 2.007 & compact\\
IRAS~10378$+$1109 &  40843 & 17.0 & 23.51 & 11.82 & 2.444 & \\
IRAS~10495$+$4424 &  27616 & 17.0 & 23.56 & 11.66 & 2.331 & \\
IRAS~10502$-$1843 &  16067 & -- & 23.01 & 11.29 & 2.470 & Sy2\\
IRAS~10565$+$2448 &  12926 & 16.0 & 23.31 & 11.56 & 2.427 & HII, interacting\\
IRAS~11010$+$4107 &  10350 & -- & 22.93 & 11.08 & 2.379 & Arp~148, ring galaxy \\
IRAS~11011$+$4345 &  14851 & -- & 23.52 & 11.22 & 1.904 & pair \\
IRAS~11069$+$2711 &  21080 & 17.0 & 23.38 & 11.23 & 2.109 & \\
IRAS~11095$-$0238 &  31796 & 17.4 & 23.75 & 11.77 & 2.154 & merger\\
IRAS~11257$+$5850 &   3101 & 12.4 & 23.14 & 11.32 & 2.327 & NGC 3690, merger\\
IRAS~11273$-$0607 &  15687 & 14.9 & 23.02 & 11.07 & 2.283 & \\
IRAS~11379$+$5338 &  27648 & -- & 23.46 & 11.46 & 2.202 & \\
IRAS~11506$+$1331 &  38206 & 17.1 & 23.64 & 11.81 & 2.355 & \\
IRAS~11529$+$8030 &  13070 & 16.8 & 23.05 & 11.09 & 2.269 & UGC 6896, quartet\\
IRAS~11554$+$1048 &  19364 & -- & 22.74 & 11.22 & 2.666 & \\
IRAS~11598$-$0112 &  45177 & 17.9 & 23.82 & 11.95 & 2.294 & X-ray source?\\
IRAS~12071$-$0444 &  38480 & 17.8 & 23.42 & 11.84 & 2.551 & Sy2\\
IRAS~12112$+$0305 &  21703 & 16.9 & 23.38 & 11.85 & 2.645 & \\
IRAS~12120$+$6838 &  18216 & 15.4 & 23.21 & 11.25 & 2.305 & \\
IRAS~12173$-$3541 &  17007 & -- & 24.24 & 11.07 & 1.081 & interacting pair\\
IRAS~12185$+$1154 &  20716 & 17.1 & 23.36 & 11.34 & 2.196 & galaxy-pair\\
IRAS~12265$+$0219 &  47469 & 13.1 & 27.42 & 11.95 & $-$1.277 & 3c273, blazar\\
IRAS~12540$+$5708 &  12518 & 14.1 & 24.02 & 11.98 & 2.099 & MRK 231, Sy1\\
IRAS~13111$+$4348 &  17290 & -- & 23.20 & 11.21 & 2.241 & \\
IRAS~13136$+$6223 &   9313 & 15.1 & 23.03 & 11.25 & 2.384 & UGC 8335, pair, HII\\
IRAS~13183$+$3423 &   6894 & 14.8 & 23.03 & 11.11 & 2.302 & IC~883, merger, LINER\\
IRAS~13225$-$2614 &  18468 & 16.2 & 22.77 & 11.15 & 2.556 & merger? starburst \\
IRAS~13333$-$1700 &  14990 & -- & 22.88 & 11.18 & 2.460 & starburst \\
IRAS~13336$-$0046 &  17859 & -- & 23.19 & 11.16 & 2.194 & UGC 8584, triple\\
IRAS~13428$+$5608 &  11180 & 15.0 & 23.59 & 11.73 & 2.282 & MRK 273, Sy2, LINER\\
IRAS~13438$-$0037 &  23058 & -- & 23.24 & 11.32 & 2.300 & \\
IRAS~13451$+$1232 &  36341 & 17.0 & 26.18 & 11.68 & $-$0.341 & Sy2, merger\\
IRAS~13454$-$2956 &  22728 & 18.6 & 23.11 & 11.34 & 2.450 & LINER\\
IRAS~14104$-$1350 &  22040 & 16.1 & 23.38 & 11.37 & 2.195 & \\
IRAS~14147$-$2248 &  23850 & -- & 23.15 & 11.41 & 2.433 & \\
IRAS~14158$+$2741 &  20902 & 16.0 & 23.37 & 11.26 & 2.118 & \\
IRAS~14290$-$2729 &  17225 & -- & 22.87 & 11.09 & 2.451 & \\
IRAS~14348$-$1447 &  24677 & 16.6 & 23.67 & 11.87 & 2.368 & LINER, merger\\
IRAS~14351$-$1954 &  26886 & -- & 23.36 & 11.48 & 2.325 & \\
IRAS~14378$-$3651 &  20277 & 16.3 & 23.49 & 11.68 & 2.354 & E?\\
IRAS~14394$+$5332 &  31450 & 17.2 & 23.96 & 11.57 & 1.776 & \\
IRAS~14547$+$2448 &  10166 & 14.6 & 23.32 & 11.08 & 2.056 & UGC 9618, galaxy pair\\
IRAS~15018$+$2417 &  20685 & 16.5 & 23.26 & 11.36 & 2.299 & \\
IRAS~15163$+$4255 &  12049 & 14.9 & 23.20 & 11.39 & 2.353 & MRK~848, HII+LINER\\
IRAS~15179$+$3956 &  14172 & 16.0 & 22.32 & 11.17 & 2.984 & merger \\
IRAS~15233$+$0533 &  16227 & -- & 22.86 & 11.24 & 2.535 & \\
IRAS~15238$-$3058 &  14318 & -- & 23.28 & 11.06 & 2.095 & \\
IRAS~15245$+$1019 &  22687 & 15.8 & 23.27 & 11.52 & 2.452 & \\
IRAS~15250$+$3609 &  16602 & 16.0 & 22.95 & 11.58 & 2.752 & ring galaxy, LINER \\
IRAS~15320$-$2601 &  20549 & -- & 22.98 & 11.40 & 2.585 & \\
IRAS~15327$+$2340 &   5534 & 14.4	& 23.33 & 11.77 & 2.607 & Arp 220, Sy2, LINER\\
IRAS~15462$-$0450 &  30150 & 16.4 & 23.41 & 11.72 & 2.459 & disturbed pair, Sy1\\
IRAS~15473$-$0520 &  16994 & -- & 22.51 & 11.17 & 2.801 & \\
IRAS~16090$-$0139 &  40029 & 16.6 & 23.87 & 12.03 & 2.344 & LINER\\
IRAS~16133$+$2107 &  27206 & -- & 23.18 & 11.48 & 2.482 & \\
IRAS~16161$+$4015 &  23288 & -- & 23.17 & 11.36 & 2.405 & \\
IRAS~16305$+$4823 &  26329 & -- & 23.04 & 11.40 & 2.523 & \\
IRAS~16429$+$1947 &  24735 & -- & 23.25 & 11.43 & 2.391 & \\
IRAS~16474$+$3430 &  33418 & 16.5 & 23.44 & 11.66 & 2.403 & \\
IRAS~16487$+$5447 &  31293 & 16.5 & 23.61 & 11.70 & 2.264 & \\
IRAS~16504$+$0228 &   7298 & 14.7 & 23.69 & 11.33 & 1.822 & NGC~6240, LINER, Sy2\\
IRAS~17028$+$5817 &  31779 & -- & 23.56 & 11.66 & 2.318 & \\
IRAS~17132$+$5313 &  15270 & -- & 23.18 & 11.42 & 2.426 & merging pair\\
IRAS~17208$-$0014 &  12836 & 15.1 & 23.46 & 12.00 & 2.703 & HII\\
IRAS~17409$+$1554 &  22042 & -- & 23.05 & 11.25 & 2.385 & \\
IRAS~17465$-$0339 &  22538 & -- & 23.30 & 11.39 & 2.245 & \\
IRAS~17487$+$5637 &  19687 & 16.9 & 22.97 & 11.27 & 2.546 & \\
IRAS~17501$+$6825 &  15357 & 15.2 & 23.12 & 11.06 & 2.194 & \\
IRAS~17517$+$6422 &  26151 & -- & 23.21 & 11.40 & 2.396 & \\
IRAS~17574$+$0629 &  32860 & 16.5 & 23.53 & 11.63 & 2.263 & E\\
IRAS~18214$-$3931 &  34005 & -- & 23.11 & 11.70 & 2.755 & \\
IRAS~18293$-$3413 &   5449 & -- & 23.16 & 11.30 & 2.344 & HII, pair\\
IRAS~18368$+$3549 &  34825 & 16.3 & 23.74 & 11.69 & 2.180 & \\
IRAS~18443$+$7433 &  40395 & 17.3 & 23.60 & 11.80 & 2.358 & \\
IRAS~18470$+$3233 &  23626 & 16.0 & 23.19 & 11.63 & 2.569 & \\
IRAS~18544$-$3718 &  22012 & -- & 22.79 & 11.40 & 2.806 & \\
IRAS~19096$+$4502 &  18980 & -- & 23.22 & 11.25 & 2.262 & pair, HII+HII\\
IRAS~19115$-$2124 &  14608 & -- & 23.45 & 11.38 & 2.156 & interacting pair\\
IRAS~19297$-$0406 &  25674 & 16.0 & 23.62 & 11.94 & 2.488 & merger, HII\\
IRAS~20010$-$2352 &  15249 & 16.9 & 22.89 & 11.19 & 2.488 & Sy2 \\
IRAS~20046$-$0623 &  25362 & 16.3 & 23.32 & 11.59 & 2.444 & interacting pair\\
IRAS~20087$-$0308 &  31600 & 16.1 & 23.11 & 11.92 & 3.009 & merger\\
IRAS~20210$+$1121 &  16905 & 15.8 & 23.52 & 11.29 & 1.873 & interacting, Sy2\\
IRAS~20414$-$1651 &  26107 & 17.1 & 23.55 & 11.76 & 2.378 & \\
IRAS~20550$+$1656 &  10900 & 15.2 & 23.05 & 11.44 & 2.523 & IIZw096, interacting\\
IRAS~20567$-$1109 &  15221 & -- & 23.26 & 11.16 & 2.068 & \\
IRAS~21316$-$1729 &  17050 & -- & 22.59 & 11.23 & 2.812 & \\
IRAS~21396$+$3623 &  29034 & 16.5 & 23.41 & 11.51 & 2.328 & \\
IRAS~21442$+$0007 &  22187 & -- & 23.10 & 11.27 & 2.413 & \\
IRAS~21504$-$0628 &  23263 & 15.6 & 23.28 & 11.56 & 2.407 & \\
IRAS~22491$-$1808 &  23312 & 16.2 & 22.84 & 11.74 & 3.025 & merger, HII \\
IRAS~23050$+$0359 &  14205 & -- & 22.85 & 11.13 & 2.474 & \\
IRAS~23204$+$0601 &  16779 & 16.0 & 23.08 & 11.33 & 2.407 & blue E\\
IRAS~23327$+$2913 &  31981 & -- & 23.27 & 11.58 & 2.493 & \\
IRAS~23365$+$3604 &  19338 & 16.3 & 23.36 & 11.71 & 2.514 & disturbed, LINER\\
IRAS~23410$+$0228 &  27335 & 17.8 & 23.02 & 11.50 & 2.608 & disburbed, X-ray source?\\
\enddata
\end{deluxetable}

\clearpage

\begin{deluxetable}{lcccccl}
\tabletypesize{\scriptsize}
\tablecaption{Radio-excess Objects \label{tab:radio-excess}}
\tablehead{
\colhead{Name}              & 
\colhead{$<cz>$}      & 
\colhead{$m$}  &
\colhead{log $L_{1.4GHz} $} &
\colhead{log $L_{60\mu m} $}      &
\colhead{q} & 
\colhead{notes} \\
\colhead{ }   &  
\colhead{km/s}   &  
\colhead{mag}   & 
\colhead{(W~Hz$^{-1}$)}   &
\colhead{$(L_\odot)$} }  
\startdata
IRAS~03164$+$4119 &  5486 & 12.4 & 25.13 & 10.57 & $-$0.463 & 3C84, NGC1275 \\
IRAS~03208$-$3723 &  1780 & 9.9 & 22.24 & 9.26 & 1.511 & NGC~1316, Fornax~A\\
IRAS~05497$-$0728 &  2334 & 14.0 & 22.54 & 9.65 & 1.301 & NGC~2110, Sy~2\\
IRAS~05511$+$4625 &  6009 & 13.9 & 23.28 & 10.32 & 1.254 & UGC3374, Sy~1 \\
IRAS~06097$+$7103 &  4050 & 13.8 & 23.59 & 10.06 & 0.618 & UGC3426, Sy~2 \\
IRAS~06488$+$2731 & 12270 & 14.9 & 23.71 & 10.83 & 1.253 & Sy~2 \\
IRAS~07224$+$3003 &  5730 & -- & 23.01 & 10.25 & 1.434 & UGC3841, FR-I RG \\
IRAS~08400$+$5023 &  3226 & 12.4 & 22.41 & 9.57 & 1.549 & NGC2639, Sy~2 \\
IRAS~10459$-$2453 &  3730 & 14.8 & 22.39 & 9.77 & 1.616 & NGC3393, Sy~2\\
IRAS~11220$+$3902 &  2080 & 12.0 & 22.02 & 9.19 & 1.535 & NGC3665, radio jet\\
IRAS~12080$+$3941 &  995 & 11.5 & 22.19 & 9.81 & 1.388 & NGC4151, Sy~1.5\\
IRAS~12173$-$3541 & 17007 & -- & 24.24 & 11.07 & 1.083 & pair \\
IRAS~12265$+$0219 & 47469 & 13.1 & 27.42 & 11.95 & $-$1.277 & 3C273, QSO\\
IRAS~13184$+$0914 &  9565 & 15.1 & 23.30 & 10.54 & 1.520 & NGC5100, pair\\
IRAS~13197$-$1627 &  5152 & 14.6 & 23.19 & 10.43 & 1.402 & Sy~2 \\
IRAS~13443$+$1439 &  6435 & 15.5 & 23.23 & 10.35 & 1.379 & Mrk~796, Sy~2\\
IRAS~13451$+$1232 & 36341 & 17.0 & 26.18 & 11.68 & $-$0.341 & 4C12.50, Sy~2\\
IRAS~13536$+$1836 & 14895 & 14.8 & 24.26 & 10.95 & 0.844 & Mrk~463, Sy~1\\
IRAS~14106$-$0258 &  1753 & 13.6 & 22.35 & 9.69 & 1.506 & NGC5506, Sy~2 \\
IRAS~15060$+$3434 & 13500 & -- & 23.72 & 10.93 & 1.475 &  VV~59, pair \\
IRAS~21497$-$0824 & 10330 & -- & 23.61 & 10.79 & 1.363 &   \\
IRAS~22045$+$0959 &  7800 & 15.1 & 23.17 & 10.51 & 1.575 & NGC7212, Sy~2 \\
IRAS~23254$+$0830 &  8698 & 13.6 & 23.55 & 10.88 & 1.530 & NGC7674, Sy~2 \\
\enddata
\end{deluxetable}


\begin{deluxetable}{lcccccl}
\tabletypesize{\scriptsize}
\tablecaption{Infrared-excess Objects \label{tab:IRexcess}}
\tablehead{
\colhead{Name}              & 
\colhead{$<cz>$}      & 
\colhead{$m$}  &
\colhead{log $L_{1.4GHz} $} &
\colhead{log $L_{60\mu m} $}      &
\colhead{q} & 
\colhead{notes} \\
\colhead{ }   &  
\colhead{km/s}   &  
\colhead{mag}   & 
\colhead{(W~Hz$^{-1}$)}   &
\colhead{$(L_\odot)$} }  
\startdata
IRAS~02244$+$4146 &  5606 & 15.0 & 21.54 & 10.35 & 3.075 & \\
IRAS~02330$-$0934 &  1509 & 11.0 & 20.36 & 9.12 & 3.085 & NGC988\\
IRAS~04332$+$0209 &  3580 & 15.6 & 21.09 & 9.96 & 3.070 & UGC3097\\
IRAS~04489$+$1029 &  8416 & -- & 21.87 & 10.47 & 3.086 & \\
IRAS~08572$+$3915 & 17480 & 14.9 & 22.50 & 11.61 & 3.218 & LINER, Sy2 \\
IRAS~12243$-$0036 &  2179 & 14.2 & 21.62 & 10.57 & 3.075 & NGC4418 \\
IRAS~19420$-$1450 &  -57 & 9.4 & 18.45 & 7.79 & 3.184 & NGC6822 \\
IRAS~20087$-$0308 & 31600 & 16.1 & 23.11 & 11.92 & 3.009 & merger\\
IRAS~22491$-$1808 & 23312 & 16.2 & 22.84 & 11.74 & 3.025 & merger, HII \\
\enddata
\end{deluxetable}

\clearpage

\begin{deluxetable}{lccccl}
\tabletypesize{\scriptsize}
\tablecaption{IRAS $60\mu m$ Luminosity Distribution and Luminosity Function \label{tab:60LF}}
\tablewidth{0pt}
\tablehead{
\colhead{ }			&
\colhead{log $L_{60\mu m}$}	&
\colhead{N}			&
\colhead{log $\rho_{60\mu m}$}  &
\colhead{$V/V_m$}   &
\colhead{ }			\\
\colhead{ }			&
\colhead{$(L_{\odot})$}		&
\colhead{ }			&
\colhead{(Mpc$^{-3}$~mag$^{-1}$)}&
\colhead{ }&  \colhead{ } }
\startdata
&7.875 &  2 & $-2.44^{+0.23}_{-0.53}$ & $0.205\pm0.320$ \\
&8.125 &  6 & $-2.34^{+0.15}_{-0.23}$ & $0.307\pm0.226$ \\
&8.375 & 16 & $-2.29^{+0.10}_{-0.12}$ & $0.449\pm0.168$ \\
&8.625 & 34 & $-2.34^{+0.07}_{-0.09}$ & $0.464\pm0.117$ \\
&8.875 & 68 & $-2.41^{+0.05}_{-0.06}$ & $0.491\pm0.085$ \\
&9.125 & 95 & $-2.64^{+0.04}_{-0.05}$ & $0.490\pm0.072$ \\
&9.375 & 128 & $-2.89^{+0.04}_{-0.04}$ & $0.411\pm0.057$ \\
&9.625 & 182 & $-3.11^{+0.03}_{-0.03}$ & $0.447\pm0.050$ \\
&9.875 & 203 & $-3.44^{+0.03}_{-0.03}$ & $0.495\pm0.049$ \\
&10.125 & 277 & $-3.68^{+0.03}_{-0.03}$ & $0.506\pm0.043$ \\
&10.375 & 242 & $-4.11^{+0.03}_{-0.03}$ & $0.507\pm0.046$ \\
&10.625 & 210 & $-4.55^{+0.03}_{-0.03}$ & $0.547\pm0.051$ \\
&10.875 & 138 & $-5.11^{+0.04}_{-0.04}$ & $0.480\pm0.059$ \\
&11.125 & 83 & $-5.70^{+0.05}_{-0.05}$ & $0.467\pm0.075$ \\
&11.375 & 57 & $-6.24^{+0.05}_{-0.06}$ & $0.534\pm0.097$ \\
&11.625 & 38 & $-6.79^{+0.07}_{-0.08}$ & $0.525\pm0.118$ \\
&11.875 & 17 & $-7.51^{+0.09}_{-0.12}$ & $0.492\pm0.170$ \\
&12.125 & 4 & $-8.52^{+0.18}_{-0.30}$ & $0.387\pm0.311$ \\
\enddata
\end{deluxetable}


\begin{deluxetable}{lcccl}
\tabletypesize{\scriptsize}
\tablecaption{Schechter Parameters for the IRAS 60 $\mu$m Luminosity 
Distribution and Function \label{tab:60LF.fits}}
\tablewidth{0pt}
\tablehead{
\colhead{Functions}	&
\colhead{$L^{*}$}	&
\colhead{$\alpha$}	&
\colhead{$\rho^*V^*$}	&
\colhead{$\chi^{2}$}	\\
\colhead{and Fits}	&
\colhead{($10^{10}L_{\odot}$)} & & or $\rho^*$ & }
\startdata
$n(L)-$low $L$ &$2.24\pm0.41$&$-0.805\pm0.058$&$509\pm 160$&$6.2$\\
$n(L)-$high $L$ &$21.46\pm1.82$&$-0.805\pm0.058$&$143\pm 8$&$4.1$\\
$n(L)-$single fit&$15.38\pm0.74$&$-1.221\pm0.019$&$309\pm 35$&$232$\\
&&&& \\
$\rho(L)-$low $L$ &$2.33\pm0.48$&$-0.821\pm0.072$&$(2.23\pm 0.29)\times10^{-4}$&$4.4$\\
$\rho(L)-$high $L$ &$21.69\pm1.85$&$-0.821\pm0.072$&$(2.21\pm 0.13)\times10^{-6}$&$4.1$\\
$\rho(L)-$single fit&$15.38\pm0.73$&$-1.220\pm0.019$&$(0.81\pm 0.07)\times10^{-5}$&$234$\\
\enddata
\end{deluxetable}

\clearpage

\begin{deluxetable}{lccccl}
\tabletypesize{\scriptsize}
\tablecaption{1.4 GHz Radio Luminosity Distribution and Luminosity Function \label{tab:1.4LF}}
\tablewidth{0pt}
\tablehead{
\colhead{ }			&
\colhead{log $L_{1.4GHz}$}	&
\colhead{N}			&
\colhead{log $\rho_{1.4GHz}$}   &
\colhead{$V/V_m$}   &
\colhead{ }                     \\
\colhead{ }			&
\colhead{(W~Hz$^{-1}$)}		&
\colhead{ }			&
\colhead{(Mpc$^{-3}$~mag$^{-1}$)}&
\colhead{ }&  \colhead{ } }
\startdata
&19.625 & 3 & $-2.64^{+0.20}_{-0.37}$ & $0.211\pm0.265$ \\
&19.875 & 8 & $-2.43^{+0.19}_{-0.34}$ & $0.284\pm0.188$ \\
&20.125 & 10 & $-2.46^{+0.14}_{-0.21}$ & $0.427\pm0.207$ \\
&20.375 & 35 & $-2.30^{+0.08}_{-0.10}$ & $0.444\pm0.113$ \\
&20.625 & 49 & $-2.49^{+0.07}_{-0.08}$ & $0.474\pm0.098$ \\
&20.875 & 66 & $-2.54^{+0.07}_{-0.08}$ & $0.428\pm0.081$ \\
&21.125 & 94 & $-2.78^{+0.05}_{-0.06}$ & $0.489\pm0.072$ \\
&21.375 & 128 & $-2.96^{+0.04}_{-0.05}$ & $0.406\pm0.056$ \\
&21.625 & 186 & $-3.11^{+0.04}_{-0.04}$ & $0.486\pm0.051$ \\
&21.875 & 228 & $-3.41^{+0.04}_{-0.04}$ & $0.495\pm0.047$ \\
&22.125 & 294 & $-3.60^{+0.05}_{-0.06}$ & $0.512\pm0.042$ \\
&22.375 & 241 & $-4.06^{+0.04}_{-0.05}$ & $0.505\pm0.046$ \\
&22.625 & 181 & $-4.52^{+0.06}_{-0.08}$ & $0.535\pm0.054$ \\
&22.875 & 115 & $-5.10^{+0.06}_{-0.07}$ & $0.465\pm0.064$ \\
&23.125 & 71 & $-5.56^{+0.09}_{-0.12}$ & $0.534\pm0.087$ \\
&23.375 & 46 & $-6.00^{+0.14}_{-0.20}$ & $0.494\pm0.104$ \\
&23.625 & 32 & $-5.95^{+0.22}_{-0.49}$ & $0.501\pm0.125$ \\
&23.875 & 4 & $-7.96^{+0.20}_{-0.37}$ & $0.753\pm0.434$ \\
&24.125 & 3 & $-7.58^{+0.28}_{-1.03}$ & $0.561\pm0.433$ \\

\enddata
\end{deluxetable}


\begin{deluxetable}{lcccl}
\tabletypesize{\scriptsize}
\tablecaption{Schechter Parameters for the 1.4 GHz Radio Luminosity 
Function \label{tab:1.4LF.fits}}
\tablewidth{0pt}
\tablehead{
\colhead{Functions}	&
\colhead{$L^{*}$}	&
\colhead{$\alpha$}	&
\colhead{$\rho^*$}	&
\colhead{$\chi^{2}$}	\\
\colhead{}	&
\colhead{(10$^{22}$ W Hz$^{-1}$)} & & & }
\startdata
$\rho(L)-$low $L$ &$2.11\pm0.28$&$-0.633\pm0.051$&$(3.17\pm 0.17)\times10^{-4}$&$5.9$\\
$\rho(L)-$High $L$ &$14.43\pm1.65$&$-0.633\pm0.051$&$(0.83\pm 0.08)\times10^{-5}$&$7.5$\\
$\rho(L)-$single fit&$10.76\pm0.84$&$-0.988\pm0.028$&$(2.43\pm 0.08)\times10^{-5}$&$112$\\
\enddata
\end{deluxetable}

\end{document}